\pdfoutput=1
\documentclass[11pt, a4paper]{article}
\usepackage[margin=2.0cm]{geometry} 
\usepackage{authblk} 

\usepackage[utf8]{inputenc} 
\usepackage{lineno}
\usepackage{hyperref} 
\hypersetup{
    colorlinks=true,             
    linkcolor={blue!80!black},   
    citecolor={blue!80!black},   
    urlcolor={blue!80!black}     
}

\usepackage{amsmath}
\usepackage{amsfonts}
\usepackage{amssymb}
\usepackage{graphicx}
\usepackage{mathrsfs}
\usepackage{times}
\usepackage{color}
\usepackage[numbers, sort&compress]{natbib} 
\usepackage{float}
\usepackage{booktabs}
\usepackage{multirow}
\usepackage{xcolor}
\usepackage{listings}
\usepackage{xurl}
\usepackage{setspace}
\usepackage{subfig}
\usepackage{enumitem}
\usepackage[font={small}]{caption}
\setlist[enumerate]{label*=\arabic*.}
\usepackage[normalem]{ulem}

\definecolor{mygreen}{rgb}{0,0.6,0}
\definecolor{myred}{rgb}{1,0,0}
\definecolor{mygray}{rgb}{0.5,0.5,0.5}
\definecolor{mymauve}{rgb}{0.88,0,0.82}
\definecolor{violet}{rgb}{0.5,0,0.5}
\definecolor{blue2}{rgb}{0.5,0,1}

\begin{document}

\title{Polymer-Resolved Multiscale Simulation of Purely Elastic Instability and Elastic Turbulence in Porous Media}

\author[1]{Edgar A. Pati\~no-Nari\~no\thanks{\texttt{epatino@bcamath.org}}}
\author[1]{Nicolas Moreno\thanks{\texttt{nmoreno@bcamath.org}}}
\author[1,2]{Marco Ellero\thanks{\texttt{mellero@bcamath.org}}}

\affil[1]{\small Basque Center for Applied Mathematics (BCAM), Alameda de Mazarredo, 14, 48009 Bilbao, Spain.}
\affil[2]{\small IKERBASQUE, Basque Foundation for Science, Calle de María Díaz de Haro 3, 48013 Bilbao, Spain.}

\date{} 

\maketitle
\vspace{-1.0cm} %

\begin{abstract}
Numerical simulation of purely elastic instabilities in complex geometries remains challenging due to strong nonlinear coupling between polymer conformation, localized stress fields, and confined flow reorganization. Here, we investigate these instabilities in periodic arrays of cylinders using the Lagrangian Heterogeneous Multiscale Method (LHMM), which couples macroscopic Smoothed Particle Hydrodynamics with microscopic Dissipative Particle Dynamics domains resolving discrete Finitely Extensible Nonlinear Elastic (FENE) bead-spring chains. By capturing bidirectional micro--macro feedback, the polymeric stress emerges naturally from chain stretching without empirical constitutive closures. For our polymeric sample, parametric evaluations identify critical instability thresholds at a Weissenberg number $\text{Wi}_c \approx 4.0$ and polymer viscosity ratio $\beta_{p,c} \approx 0.90$. Above onset, flow symmetry breaking initiates self-sustained, ``whip-like'' transverse velocity oscillations driven by intense molecular stretching at trailing stagnation points, generating localized bands of first normal stress difference and shear stress. Downstream advection and interaction of unrelaxed elastic wakes across successive obstacles trigger a cooperative ``mixer effect,'' producing broadband velocity spectra with apparent power-law slopes ($\beta \approx 3\text{--}4$), consistent with elastic-turbulence-like scaling. Varying array porosity ($\varepsilon$) reveals a transition from elastic stresses pinned in narrow constriction gaps ($\varepsilon \le 0.80$) to mobile, elongated ``birefringent strands'' acting as dynamic flow barriers ($\varepsilon = 0.91$). Tracking the in-plane gyration tensor ($\mathbf{R}_g$) shows that intense microstructural deformation and strong broadening of the conformational distribution away from the equilibrium Gaussian precede the macroscopic unsteady velocity response with a measurable temporal offset, linking these viscoelastic regimes to early-stage microstructural changes.
\end{abstract}

\section{Introduction} \label{sec:intro}
At vanishing Reynolds numbers ($\text{Re}\ll1$), viscoelastic flows through porous media can undergo a sharp transition from a steady state to a nonlinear unsteady regime. Under complex geometric confinement, this transition is driven by nonlinear elastic stresses that develop along curvilinear streamlines, a phenomenon widely recognized as purely elastic instability (EI) \cite{GroismanA.2000, Steinberg2021}. While foundational studies on EI were primarily conducted using constant-viscosity Boger fluids to isolate elastic effects, recent evidence \cite{Jha2021, DeBlois2023, Machado2016, Chen2025} shows that for strongly shear-thinning semidilute polymer solutions, the onset and scaling of these instabilities are further influenced by their shear-dependent rheological properties, leading to a flow resistance that significantly exceeds Newtonian predictions \cite{Varshney2018a, Garg2018, CastilloSanchez2022, Shakeri2022}.

In the context of porous-like media, the Periodic Array of Cylinders (PAC) serves as a canonical geometry to study these effects. Research by Ichikawa et al. \cite{Ichikawa2022,Ichikawa2025}, Datta et al. \cite{Datta2022, Chen2025}, Mokhtari et al. \cite{Mokhtari2022,Mokhtari2024}, and Zhu et al. \cite{Zhu2022a, Zhu2024} has highlighted how inter-obstacle spacing and stagnation points act as catalysts for elastic stress accumulation. In particular, the experimental work of Haward et al. \cite{Haward2019, Haward2021c,Recktenwald2025} and Ichikawa \& Motosuke \cite{Ichikawa2022,Ichikawa2025} has shown the central role of the cylinder's stagnation points (``dead zone'') in generating ``birefringent strands,'' localized regions of high polymer extension originating at the trailing edge of the cylinder. These strands form the backbone of the elastic wake and are the primary drivers of the symmetry-breaking transitions observed in confined geometries \cite{Beech2023,Turato2026,Beneitez2023}.

The formation of these structures is driven by a nonlinear feedback loop between the flow field and polymer conformation \cite{Browne2020, Guimaraes2023, Recktenwald2025}. Theoretical and numerical studies \cite{Morozov2022, Lellep2023} show that this mechanism operates via a self-amplifying coupling: velocity fluctuations stretch the polymers, generating localized elastic stress gradients that feed back into the velocity field. In a PAC, geometric confinement intensifies this feedback, as the elastic wake of a leading cylinder acts as a finite-amplitude perturbation for the downstream stagnation zone. This cooperative interaction lowers the instability threshold and gives rise to \textit{Exact Coherent Structures} (ECS), including localized polymer tension sheets and arrowhead-like conformation gradients, which propagate through the lattice as traveling waves \cite{Page2020, Lellep2023}.

Recent studies by Mokhtari et al. \cite{Mokhtari2024} suggest that these strands behave as a ``sticky web'' of internal flow barriers. These self-organizing stress barriers actively reorganize the flow, forcing the fluid into preferential paths and creating states of multistability. In this regime, the global transport pattern becomes history-dependent, where the fluid is preferentially routed along competing stable pathways based on the local Weissenberg number ($\text{Wi}$) and the degree of geometric confinement imposed by the lattice \cite{Zhu2024, Beneitez2024}. In rectilinear viscoelastic channel flows, the transition from coherent motion to complex dynamics is mediated by center-mode instabilities \cite{Khalid2021a} and finite-amplitude elastic waves \cite{Buza2022a}. In a PAC, however, resolving the polymeric stress fields directly from the microscale allows us to examine how curvature-driven elastic waves and coherent stress barriers govern this transition in a shear-thinning polymer system.

This route from steady-laminar states to a self-sustained chaotic regime is critically sensitive to the pore fraction (porosity) and microscale polymer dynamics \cite{Hopkins2022, Haward2019}. Rather than appearing abruptly, the transition develops through the growth and destabilization of localized birefringent strands. These structures are persistently anchored near stagnation points even in steady flows, mirroring the highly localized polymeric stress sheets detailed in the frameworks of Berti and Boffetta \cite{Berti2008,Berti2010}. As the flow destabilizes, they evolve into highly unsteady, fluctuating networks whose multiscale tracking remains computationally demanding \cite{King2025,Morozov2007}. In multi-obstacle environments, downstream advection of unrelaxed elastic wakes triggers a cooperative ``mixer effect,'' where wake interference acts as a finite-amplitude perturbation that lowers the critical threshold and drives the transition toward spatiotemporal elastic turbulence with characteristic power-law spectral decay ($\beta \approx 3\text{--}4$) \cite{FoggiRota2026, DeBlois2023, Pan2013}. Furthermore, this route to chaos is closely linked to the exponential growth of microscopic conformational uncertainty \cite{King2025}, often initiated by specific bifurcation sequences like period-doubling \cite{Nichols2025a} or wall-diffusive instabilities \cite{Beneitez2023}.

Traditional macroscopic continuum solvers face well-known difficulties in capturing these transitions without introducing artificial regularizations \cite{Dubief2023, Alves2021}. Standard models (such as Oldroyd-B or FENE-P) rely on closed constitutive equations that inherently suffer from stress singularities at stagnation points under high $\text{Wi}$ \cite{Shaqfeh2021, Mokhtari2022, Datta2022}. To remain numerically stable, these solvers often require artificial stress diffusion or rely on constitutive approximations (such as the Peterlin spring-closure in FENE-P) \cite{Sureshkumar1995, Thomases2011, Hu2026}. Consequently, they artificially damp extreme stress gradients and obscure the true microscopic memory and fluctuation dynamics inherent to highly localized polymeric sheets \cite{Mokhtari2022, Page2020, Mokhtari2023, Zhu2025}.

To address these limitations, this work employs a constitutive-free multiscale approach that captures purely elastic instabilities directly from a microscopic perspective. We adopt the Lagrangian Heterogeneous Multiscale Method (LHMM) \cite{moreno2023, Patino-Narino2025a, Moreno2026a}, where a thermodynamically consistent Smoothed Particle Hydrodynamics (SPH) macroscale \cite{Ellero2018, Vazquez-Quesada2015} is bidirectionally coupled to mesoscopic Dissipative Particle Dynamics (DPD) subdomains containing discrete Finitely Extensible Nonlinear Elastic (FENE) polymer chains \cite{Fedosov2010, Moreno2015a, Moreno2014b}. By maintaining a fully Lagrangian representation along material streamlines, local macroscopic velocity gradients imposed via Arbitrary Flow Boundary Condition (AFBC) \cite{Moreno2021} drive the deformation of discrete polymer ensembles. The resulting Irving--Kirkwood stress tensor feeds back on-the-fly into the continuum momentum balance \cite{Patino-Narino2025a}. This bottom-up formulation explicitly models collective polymer behavior at the microscale, allowing shear-thinning rheology, normal stress differences, and history-dependent viscoelastic relaxation to emerge naturally from the physics of the chains without relying on empirical macroscopic closures.

Resolving the polymer microstructure explicitly gives direct access to chain-level dynamics that are damped or absent in continuum closures. In addition to capturing macroscopic instabilities, the LHMM framework resolves the microstructural deformation behind the sticky flow barriers (birefringent strands) at trailing stagnation points. We demonstrate how the localized stretching of discrete polymer chains reorganizes macroscopic transport pathways and drives lateral symmetry breaking. Furthermore, our simulations link the cooperative mixer effect observed in multi-obstacle arrays directly to incomplete chain relaxation between obstacles. We show how the downstream advection of unrelaxed elastic wakes drives the flow toward an elastic-turbulence-like regime with broadband spectral decay ($\beta \approx 3\text{--}4$) \cite{FoggiRota2026, DeBlois2023, Pan2013}. By tracking the in-plane gyration tensor of the discrete bead-spring chains and their local non-equilibrium probability density functions (PDFs), we show that the physical reorientation and extension of the polymers systematically precede macroscopic kinematic symmetry breaking. Statistically, the onset of instability appears as a marked broadening in the stretching variance of the chains across the birefringent strands, reflecting a highly heterogeneous local state where strongly extended and coiled polymers coexist within the same micro-domain.

The remainder of this paper is structured as follows. Section~\ref{sec:numerical_method} details the multiscale LHMM framework, outlining the Lagrangian coupling strategy between the SPH macroscale \cite{Patino-Narino2023, Patino-Narino2025} and the DPD-FENE microscale \cite{Moreno2020, Moreno2025, Moreno2015}. Section~\ref{sec:results} analyzes the emerging flow dynamics, beginning with the macroscopic stress evolution and the onset of EI at critical thresholds of $\text{Wi}_c \approx 4.0$ and $\beta_{p,c} \approx 0.90$, followed by the effects of wake interference in multi-obstacle arrays and array porosity ($\varepsilon$). We conclude Section~\ref{sec:results} with a microstructural analysis linking normal and shear stress structures to the physical deformation of the underlying polymer chains via 2D tensor glyphs and local conformational PDFs, showing the temporal precedence of chain reorganization and conformational broadening away from the equilibrium Gaussian over transverse flow deflection. Finally, Section~\ref{sec:conclusions} summarizes our main findings and outlines future directions for the LHMM in complex porous media applications.

\section{Numerical Method and Modeling}\label{sec:numerical_method}

\begin{figure}[!ht]
    \centering
    \includegraphics[width=\columnwidth]{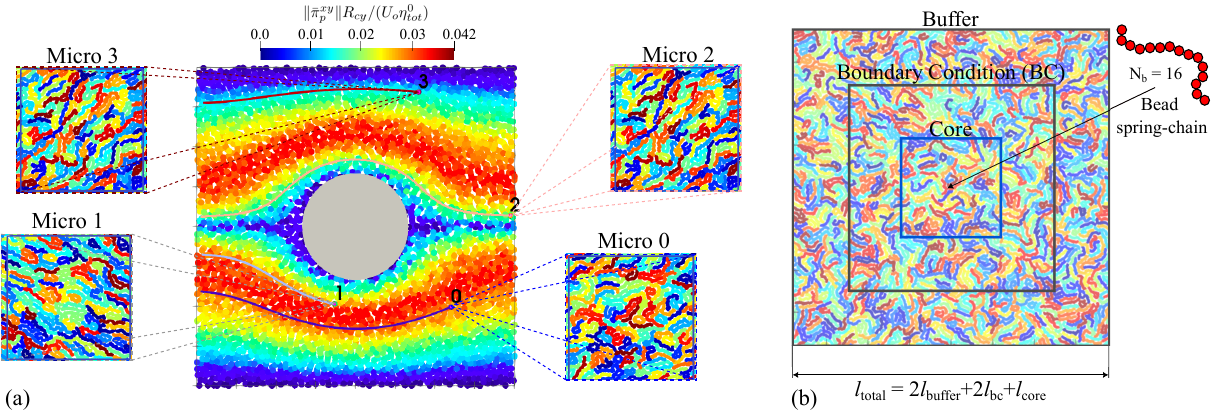}
    \caption{Multiscale LHMM framework for PAC flow. The illustrated case corresponds to a representative steady-state response. 
    (a) Micro-macro mapping: Spatial distribution of the normalized polymeric shear stress. SPH particles are colored by their local stress contribution. Four representative particles (0, 1, 2, and 3) show their Lagrangian trajectories and the corresponding snapshots of their microscopic DPD domains, revealing the instantaneous stretching of the FENE chains.
    (b) AFBC: Schematic of the microscopic domain $\Omega'$, partitioned into the buffer ($\Omega'_0$), boundary condition ($\Omega'_{\rm bc}$), and core ($\Omega'_{\rm core}$) regions. The lengths $l_{\rm total}$, $l_{\rm  core}$, $l_{\rm bc}$, and $l_{\rm  buffer}$ define the spatial decomposition for consistent top-down velocity imposition.}
    \label{Fig:1}
\end{figure}

The multiscale architecture of the LHMM framework \cite{moreno2023}, which bridges mesoscopic polymer dynamics and continuum fluid mechanics, is summarized in Fig.~\ref{Fig:1}. This methodology couples the scales bidirectionally, such that the macroscopic flow field is locally informed by the microscopic stress response of suspended DPD/FENE chains. As illustrated in Fig.~\ref{Fig:1}(a), this is exemplified by the color-coded SPH particles representing the polymeric shear stress component $\bar{\pi}_p^{xy}$ of the tensor $\bar{\boldsymbol{\pi}}_p$.  To visualize the micro-to-macro mapping in Fig.~\ref{Fig:1}(a), four representative Lagrangian SPH particles (labeled $I=0, 1, 2, 3$) are tracked along their streamlines as they navigate the PAC domain. Their corresponding microscopic snapshots reveal the instantaneous spatial organization and stretching of the bead chains within their local core domains. 

Meanwhile, Fig.~\ref{Fig:1}(b) details the internal structure of these microscopic domains $\Omega'$ using the AFBC protocol~\cite{Moreno2021}. The domain is partitioned into buffer ($\Omega'_0$), boundary ($\Omega'_{\rm bc}$), and core ($\Omega'_{\rm core}$) regions, as described in Appendix~\ref{sec:boundary_regions}. Following the nomenclature in \cite{moreno2023,Patino-Narino2025a}, we adopt a consistent notation: variables associated with macroscale fluid particles use uppercase indices ($I, J$), while microscale beads use lowercase indices ($i, j$). Moreover, statistical-mechanical projections from the micro-domain $\Omega'$ into the macroscale are denoted with an overbar (e.g., the polymeric stress tensor $\bar{\boldsymbol{\pi}}_{p} = \langle \boldsymbol{\pi}'_p \rangle$), whereas variables defined strictly within the micro-domain use prime notation. In the remainder of this section, we summarize the LHMM methodology; full details on the underlying theoretical formulation and solid-boundary (no-slip) treatment for stationary obstacles are provided in the original framework~\cite{moreno2023} and our previous implementation study~\cite{Patino-Narino2025a}.

\subsection{LHMM Framework and Governing Equations}\label{LHMM}
The LHMM couples continuum mechanics with explicit microscale dynamics \cite{moreno2023}. At the macroscale, an incompressible viscoelastic fluid is governed by the conservation of mass and momentum:
\begin{equation}
    \nabla \cdot \mathbf{v}=0, \quad 
    \rho\frac{d\mathbf{v}}{dt}=\nabla\cdot\boldsymbol{\sigma}+\mathbf{f},
    \label{eq:NS_01} 
\end{equation}
where the total stress is $\boldsymbol{\sigma} = -p\mathbf{I} + \boldsymbol{\tau}_s + \bar{\boldsymbol{\pi}}_p$. Here, $\boldsymbol{\tau}_s = 2\eta_s\mathbf{d}$ is the Newtonian solvent stress (with $\eta_s$ being the solvent viscosity and $\mathbf{d}$ the strain-rate tensor), and $\bar{\boldsymbol{\pi}}_p$ is the polymeric stress tensor derived from the microscale dynamics~\cite{Patino-Narino2025a}. The term $\mathbf{f}$ represents an external body force vector; in the present configuration, it is defined as a constant driving force acting in the $x$-direction, $\mathbf{f} = (F, 0)$. 

\subsection{Macroscale: Smoothed Particle Hydrodynamics}
The macroscopic momentum equation is discretized into a set of $N_{\text{macro}}$ Lagrangian particles, indexed by $I = 1, \dots, N_{\text{macro}}$, following a thermodynamically consistent SPH formulation \cite{Espanol2003, Ellero2018}. The momentum equation for a particle $I$ interacting with its neighbors $J \in \{1, \dots, N_{\text{macro}}\}$ is given by:

\begin{equation} \label{eq:sph_momentum}
    m_I\frac{d\mathbf{v}_I}{dt} = \sum_J \left[\frac{p_I}{d_I^2} + \frac{p_J}{d_J^2}\right] F_{IJ}\mathbf{r}_{IJ}
    - \sum_J \left[a\mathbf{v}_{IJ} + b(\mathbf{v}_{IJ}\cdot\mathbf{e}_{IJ})\mathbf{e}_{IJ}\right]\frac{F_{IJ}}{d_I d_J}
    - \sum_J \bar{\boldsymbol{\pi}}_{p,IJ} F_{IJ}\mathbf{r}_{IJ}
\end{equation}
where $\mathbf{r}_{IJ} = \mathbf{r}_I - \mathbf{r}_J$, $\mathbf{v}_{IJ} = \mathbf{v}_I - \mathbf{v}_J$, and $\mathbf{e}_{IJ} = \mathbf{r}_{IJ}/\|\mathbf{r}_{IJ}\|$ is the unit vector between particles. The particle number density is defined as $d_I = \sum_{J} W(\|\mathbf{r}_{IJ}\|, h)$, and the scalar kernel gradient term is $F_{IJ} = -\nabla W_{IJ}/\|\mathbf{r}_{IJ}\|$. We adopt a Lucy kernel, $W_{IJ} = W^{\text{Lucy}}(\|\mathbf{r}_{IJ}\|, h)$, with a smoothing length of $h=4\Delta x$, and $\Delta x$ being the mean SPH particle spacing. The friction coefficients $a = 2\eta_s - \zeta_s$ and $b = 4\zeta_s + 2\eta_s$ relate the particle interactions to the solvent shear ($\eta_s$) and bulk ($\zeta_s$) viscosities \cite{moreno2023, Ellero2018}. The pressure $p_I$ is governed by the Tait equation of state:

\begin{equation} \label{eq:tait}
    p_I(\rho_I) = \frac{c^2 \rho_0}{7} \left[ \left(\frac{\rho_I}{\rho_0}\right)^7 - 1 \right] + p_b
\end{equation}

where $\rho_I = m_I d_I$ is the mass density, $c$ is the numerical speed of sound, and $p_b$ is a background pressure introduced to prevent the pressure field from becoming negative and to avoid tensile instability. To ensure numerical stability while maintaining weak compressibility, the Mach number is restricted to $\text{Ma} = U_0/c \leq 0.1$, where $U_0$ is the average characteristic velocity of the fluid.

\subsection{Microscale: Polymer Melt (DPD/FENE)} \label{subsec:micro}

Microscopic domains utilize DPD to model polymer chains as bead-spring systems. Each microscopic bead $i \in \{1, \dots, N_{\text{micro}}\}$ within the micro-domain $\Omega'_I$ associated with macro-particle $I$ evolves according to the following equations of motion:
\begin{equation}
m_i \frac{d\mathbf{v}'_i}{dt} = \sum_{j \neq i}^{N_{\text{micro}}} \left( \mathbf{F}_{ij}^C + \mathbf{F}_{ij}^D + \mathbf{F}_{ij}^R \right) + \mathbf{F}_{i}^{\text{bond}}
\label{eq:motion}
\end{equation}
where $\mathbf{F}_{ij}^C, \mathbf{F}_{ij}^D, \text{ and } \mathbf{F}_{ij}^R$ represent the conservative, dissipative, and random forces, respectively, acting between beads $i$ and $j$ \cite{Espanol1995}. Chain connectivity is maintained via the FENE potential, which contributes to the bonding force $\mathbf{F}_{i}^{\text{bond}}$~\cite{Fedosov2010, Litvinov2014}. The force between two bonded beads is given by: $\mathbf{F}_{ij}^{\text{FENE}} = -\mathbf{r}'_{ij}H_{\text{FENE}}/ [1 - (\|\mathbf{r}'_{ij}\|/R_0)^2]$, where $H_{\text{FENE}}$ is the spring constant, $\mathbf{r}'_{ij}$ defines the relative distance between beads $i$ and $j$, and $R_0$ is the maximum extensibility. A detailed description of the DPD force parameters and the non-dimensionalization procedure is provided in Appendix \ref{app:DPD_details}, with core parameters summarized in Table \ref{Tab:1}. For each macro-particle, an ensemble of chains (with $N_b=16$ beads per chain) is simulated to compute the Irving–Kirkwood (IK) stress $\langle\boldsymbol{\pi}'_{p}\rangle$, as derived in Appendix \ref{sub_micro_IK}.

\begin{table}[ht]
    \centering
    \caption{DPD and FENE parameters for the polymer microsystems in reduced units.}
    \label{Tab:1}
    \begin{tabular}{cccccccccc}
    \hline 
    $k_{B}T$ & $a_{ij}$ & $\gamma^D$ & $m_{\text{DPD}}$ & $n_{\text{DPD}}$ & $\Delta x'$ & $R_{0}$ & $r_{c}$ & $H_{\text{FENE}}$ & $\Delta t_{\text{DPD}}$ \\ \hline 
    1.0 & 25 & 4.5 & 1 & 4 & 0.5 & 1.5 & 1.0 & 50 & 0.005 \\ \hline 
    \end{tabular}
\end{table}

\subsection{Bidirectional Coupling and Synchronization}\label{Pi_p}
The LHMM utilizes a bidirectional data exchange to bridge the scale gap. \textit{Top-down:} The macroscale velocity gradient $\nabla \mathbf{v}_I$ is computed locally via the SPH kernel approximation:
\begin{equation} \label{eq:macro_micro}
    \nabla \mathbf{v}_I = \sum_J F_{IJ} \mathbf{r}_{IJ} \mathbf{v}_{IJ}.
\end{equation}
This gradient is imposed on the microscopic domain $\Omega'_I$ associated with each macro-particle using the AFBC protocol~\cite{Moreno2021}. Although the macroscopic ($\bar{\lambda}$) and microscopic ($\lambda'$) relaxation times could theoretically differ, they are set to unity ($\lambda^* = \bar{\lambda}/\lambda' = 1$) by identifying both with the characteristic Carreau--Yasuda relaxation time $\lambda$ (Sec.~\ref{sec:micro_results}). Because the macroscopic rheology emerges directly from the microscopic chain dynamics \cite{Patino-Narino2025a}, this choice ensures dynamical consistency and simplifies the coupling: the stress magnitude is matched via the scaling factor $\alpha$ in Eq.~\eqref{eq:PI_P}, avoiding additional time-scale corrections.

\textit{Bottom-up:} The microscopic IK stress is statistically projected and scaled to close the macroscopic momentum equation. The polymeric contribution in Eq.~\eqref{eq:sph_momentum} is given by:
\begin{equation}\label{eq:PI_P}
    \bar{\boldsymbol{\pi}}_{p,IJ} = \alpha \left( \frac{\langle\boldsymbol{\pi}'_{p}\rangle_I}{d_I^2} + \frac{\langle\boldsymbol{\pi}'_{p}\rangle_J}{d_J^2} \right),
\end{equation}
where $\alpha = \bar{\eta}_p^0 / \eta_p'^0$ is the viscosity scaling factor. To maintain consistency across scales, the reference microscopic viscosity is set to $\eta_p'^0 = \eta_{\text{CY}}^0$, the zero-shear viscosity obtained from the microscale characterization (Sec.~\ref{sec:micro_results}). Within this splitting scheme, fast Newtonian solvent dissipation ($\eta_s$) is resolved directly at the continuum SPH level via the viscous term in Eq.~\eqref{eq:sph_momentum}, while the solvent-free DPD micro-domains evaluate the configurational polymeric stress. The factor $\alpha$ thus bridges both scales so that the macroscopic polymeric viscosity $\bar{\eta}_p^0 = \alpha \eta_{\mathrm{CY}}^0$ satisfies the target viscosity ratio $\beta_p = \bar{\eta}_p^0 / \eta_{\mathrm{tot}}^0 = \alpha \eta_{\mathrm{CY}}^0 / (\alpha \eta_{\mathrm{CY}}^0 + \eta_s)$, with $\eta_{\mathrm{tot}}^0 = \bar{\eta}_p^0 + \eta_s$.

\subsection{Rheological Characterization} \label{sec:micro_results}

\begin{figure}[!ht]
    \centering
    \includegraphics[width=\columnwidth]{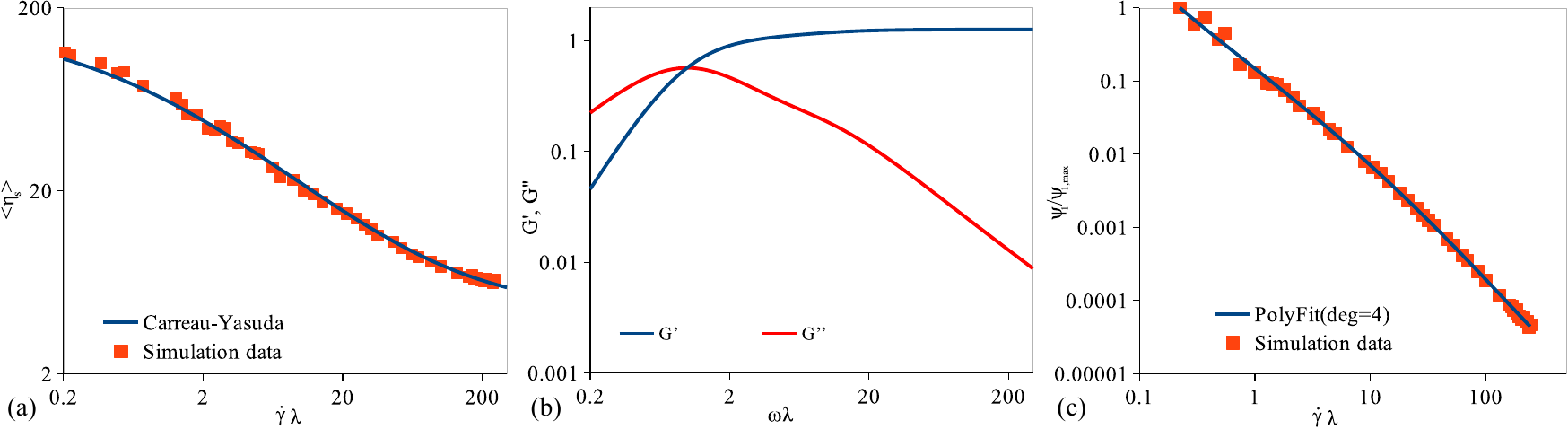}
    \caption{Microscopic rheological characterization of the $N_b=16$ DPD polymer melt \cite{Patino-Narino2025a}. (a) Steady-state shear viscosity as a function of shear rate $\dot{\gamma}$; symbols denote DPD simulation data, while the solid line represents the Carreau--Yasuda fit. (b) Dynamic moduli, storage ($G'$) and loss ($G''$), as a function of angular frequency $\omega$, obtained via Inverse Laplace Transform (ILT). (c) First normal stress coefficient ($\Psi_1$) under steady shear flow; the solid line represents the degree-four polynomial fit (PolyFit), and $\Psi_{1,\text{max}}$ denotes the maximum value of $\Psi_1$.}
    \label{fig:rheology_summary}
\end{figure}

The rheological behavior of the $N_b=16$ polymer melt was characterized using a virtual rheometer to bridge the microscopic DPD dynamics with the macroscopic framework \cite{Patino-Narino2025a}. The steady-state shear viscosity $\eta(\dot{\gamma}) = \bar{\pi}^{xy}_{p}/\dot{\gamma}$ was extracted (Fig.~\ref{fig:rheology_summary}(a)) and fitted to the empirical Carreau--Yasuda (CY) model:
\begin{equation} \label{eq:Yasuda}
    \eta(\dot{\gamma}) = \eta_{\text{CY}}^\infty + (\eta_{\text{CY}}^{0} - \eta_{\text{CY}}^\infty) \left[ 1 + (\lambda \dot{\gamma})^{a_{\text{CY}}} \right]^{\frac{n_{\text{CY}}-1}{a_{\text{CY}}}},
\end{equation}
where $\dot{\gamma}$ is the shear rate, $\eta_{\text{CY}}^{0} = 173.18$ and $\eta_{\text{CY}}^\infty = 4.45$ are the zero-shear and infinite-shear viscosities, $\lambda = 353.48$ is the characteristic polymeric relaxation time, $a_{\text{CY}} = 0.55$ controls the transition region, and $n_{\text{CY}} = 0.18$ is the power-law exponent (all in reduced DPD units; see Appendix~\ref{app:DPD_details} for detailed DPD parameters). Furthermore, the linear viscoelastic spectrum and the first normal stress coefficient, defined as $\Psi_1 = (\bar{\pi}_{p}^{xx} - \bar{\pi}_{p}^{yy}) / \dot{\gamma}^2$, are shown in Figs.~\ref{fig:rheology_summary}(b) and (c), respectively.


\subsection{Computational Implementation}
The framework is implemented in a hybrid CPU-GPU LAMMPS engine \cite{Patino-Narino2025a}. The architecture employs MPI sub-communicators to manage independent micro-domains, utilizing a GPU-enabled \texttt{DPD/GPU} module for microscopic integrations. This C++/MPI design minimizes communication overhead, achieving strong scaling for systems with up to $10^8$ total particles. The source code is available at \url{https://github.com/BCAM-CFD/LHMM}. Furthermore, the specific details regarding numerical stability and scaling constraints are discussed in Appendix~\ref{app:stability_scaling}.

\section{Results and Discussion}\label{sec:results}
\begin{figure}[!ht]
    \centering
    \includegraphics[scale=0.8]{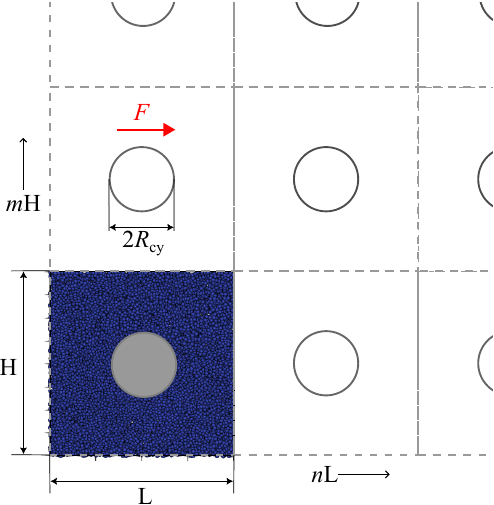}
    \caption{Computational PAC configuration: Definition of the $n\text{L} \times m\text{H}$ array. The unit cell consists of a square domain with a centered cylindrical obstacle of radius $R_{\text{cy}}$. Periodic boundary conditions are imposed in $x$ and $y$, with a constant body force $F$ driving the flow.}
    \label{Fig:1Geo}
\end{figure}

To describe the transport within the PAC geometry, we define the baseline configuration shown in Fig.~\ref{Fig:1Geo}, consisting of an $n\text{L} \times m\text{H}$ computational domain with periodic boundary conditions in both $x$ and $y$. While periodicity in a single unit cell already represents an infinite lattice, $n$ and $m$ denote the number of cylinders modeled explicitly in the streamwise and transverse directions; any differences observed as $n$ and $m$ are varied therefore reflect periodic finite-size effects that constrain wake decorrelation in smaller domains. Following our previous work \cite{Patino-Narino2025a}, the flow is characterized by the polymer viscosity ratio $\beta_p$, the Weissenberg number $\text{Wi} = U_0 \lambda / R_{\text{cy}}$, and the Reynolds number $\text{Re} = R_{\text{cy}} U_0 \rho / \eta_{\text{tot}}^0$, where $U_0$ represents the average fluid velocity. All temporal evolutions are scaled by the polymer relaxation time $\lambda$, ensuring a consistent mapping between micro- and macroscales. 

Simulations were conducted for the PAC configurations over a wide range of Wi at low Re. The flow is driven by distinct body force magnitudes $F$, which determine the characteristic velocity $U_0$. This setup allows us to explore a parameter space spanning $1.0 \times 10^{-3} < \text{Re} < 0.05$ and $1.0 < \text{Wi} < 130$, with a primary focus on the high-concentration regime ($\beta_p = 0.97$, $\eta_s = 0.08$). Spatial convergence was verified in Appendix~\ref{sec:results_1PAC_PSD}, establishing $R_{\text{cy}}/\Delta x = 7.0$ as the reference resolution. Specifically, as detailed in Appendix~\ref{sec:scalabilityEI}, a $3\text{L} \times \text{H}$ array ($\text{L} = \text{H} = 4$, $R_{\text{cy}} = 0.56$, yielding a porosity of $\varepsilon = 1 - \phi = 1 - \pi R_{\text{cy}}^2 / (\text{LH}) = 0.94$, where $\phi$ is the solid volume fraction; Fig.~\ref{Fig:1Geo}) is adopted as our primary operational benchmark. This configuration is chosen because an array of at least three obstacles ($n \ge 3$) represents the shortest tested longitudinal domain for which the spectral response becomes nearly unchanged, indicating that the dominant wake-interaction dynamics are captured at this domain length and that the PSD approaches a converged broadband decay ($\beta \approx 3\text{--}4$), reducing the constraints of smaller periodic cells without excessive computational cost.

\subsection{Physical Mechanism of Purely Elastic Instability (EI)}
\begin{figure}[!ht]
   \centering
   \includegraphics[width=\columnwidth]{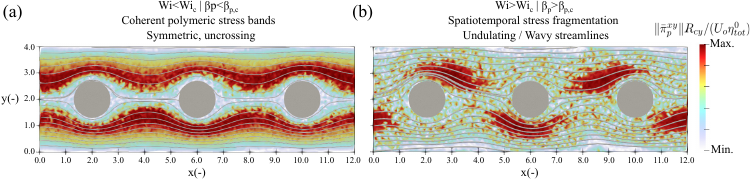}
   \caption{Schematic representation of the onset of purely elastic instability across a cylinder array. Spatial distribution of the normalized polymeric shear stress $\|\bar{\pi}_p^{xy}\| R_{\text{cy}} / (U_0 \eta_{\text{tot}}^0)$ overlaid with fluid streamlines. (a) Below the critical threshold ($\text{Wi} < \text{Wi}_c, \beta_p < \beta_{p,c}$): Steady, spatially symmetric flow characterized by connected, coherent polymeric stress bands channeled through the constriction gaps. (b) Above the critical threshold ($\text{Wi} > \text{Wi}_c, \beta_p > \beta_{p,c}$): Purely elastic symmetry breaking, marked by the spatiotemporal fragmentation of the polymeric stress field into localized pockets and undulating, asymmetric streamlines.}
   \label{fig:schematic_onset}
\end{figure}

A primary feature of EI at $\text{Re} \ll 1$ in confined geometries is the transition from a spatially symmetric flow to a symmetry-broken, spatiotemporally dynamic state \cite{Poole2007, Grilli2013, Mokhtari2022,Takizawa2026}. This transition is governed by the interplay between fluid elasticity ($\text{Wi}$) and polymer contribution or viscosity ratio ($\beta_p$) \cite{Shakeri2022,Mohanty2024,Chen2025,Turato2026}. As schematically illustrated in Fig.~\ref{fig:schematic_onset}, below the critical threshold ($\text{Wi} < \text{Wi}_c, \beta_p < \beta_{p,c}$), the microscopic polymer stress $\bar{\pi}_p^{xy}$ self-organizes into a continuous, spatially coherent stress field coupled with steady-state streamlines (Fig.~\ref{fig:schematic_onset}(a)), in agreement with  stable steady-state stress filaments past obstacles \cite{Khalid2021, Mokhtari2022, Morozov2022}. Exceeding this boundary triggers the instability, driving the spatiotemporal fragmentation of these coherent stress structures into isolated, alternating high-stress pockets accompanied by undulating, asymmetric streamlines (Fig.~\ref{fig:schematic_onset}(b)) \cite{Mokhtari2024, Peng2023}. In the following sections, we systematically explore this stability phase space $(\text{Wi}, \beta_p)$ using our multiscale framework.

\subsection{Micro-Macro Stress Evolution: Normal and Shear Stress Dynamics}\label{sec:results_stress}
To identify the driving forces of the instability, we examine the evolution of the first normal stress difference $\bar{N}_{1} = \bar{\pi}^{xx}_p - \bar{\pi}^{yy}_p$ and the shear stress magnitude $\|\bar{\pi}^{xy}_p\|$, derived directly from the microscale polymeric stress tensor $\bar{\boldsymbol{\pi}}_p$, for a representative unstable state ($\text{Wi}=91.7$, $\text{Re}=0.035$, $F=0.02$).

\subsubsection{Transient Stress Accumulation and Wake-Obstacle Breakdown}
\begin{figure*}[!ht]
    \centering
    \includegraphics[width=\textwidth]{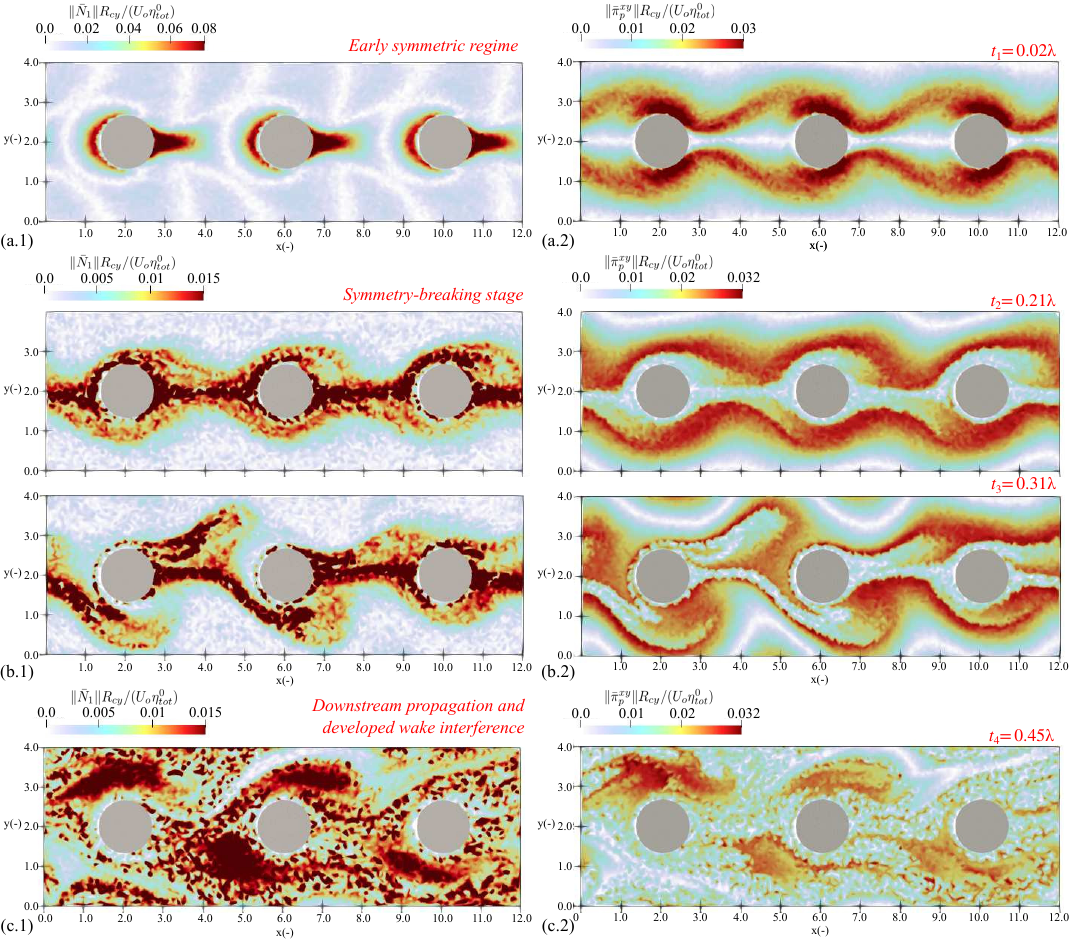}
    \caption{Spatiotemporal transient evolution of the elastic instability in a $3\text{L} \times \text{H}$ multi-obstacle array ($F=0.02$, $R_{\text{cy}}/\Delta x = 7.0$). Panels display sequential snapshots of the dimensionless first normal stress $\|\bar{N}_1\| R_{\text{cy}} / (U_0 \eta_{\text{tot}}^0)$ (indices .1) and shear stress $\|\bar{\pi}^{xy}_p\| R_{\text{cy}} / (U_0 \eta_{\text{tot}}^0)$ (indices .2): 
    (a) Early symmetric regime ($t_1 = 0.02\lambda$), showing localized stress accumulation pinned at stagnation points. 
    (b) Symmetry-breaking stage shown at two distinct instances ($t_2 = 0.21\lambda$ and $t_3 = 0.31\lambda$), where localized elastic saturation triggers initial wake deflection behind the leading cylinder. 
    (c) Downstream propagation and developed wake interference ($t_4 = 0.45\lambda$), illustrating the spatiotemporal progression of the mixer effect across the lattice cells.}
    \label{fig:transient_evolution}
\end{figure*}

During early start-up ($t < 0.05\lambda$), the polymer ensemble undergoes longitudinal stretching at the extensional stagnation points, producing a symmetric accumulation of $\bar{N}_1$ alongside a complementary $\|\bar{\pi}_{p}^{xy}\|$ distribution across all lattice units (Figs.~\ref{fig:transient_evolution}(a.1) and \ref{fig:transient_evolution}(a.2)) \cite{Patino-Narino2025a}. Once elastic stress along the cylinder boundaries reaches saturation ($t \approx 0.21\lambda$), this symmetric state destabilizes and sheds elastic tension sheets downstream (Fig.~\ref{fig:transient_evolution}(b)) \cite{Mokhtari2024}.

The elastic wake shed by upstream cylinders acts as a finite-amplitude perturbation that bypasses the linear stability threshold of downstream obstacles \cite{Pan2013, Varshney2018}. This wake--obstacle interference initiates the cooperative ``mixer effect'' (Fig.~\ref{fig:transient_evolution}(c)), where the advection of unrelaxed elastic stresses disrupts local phase-locking well before the flow reaches its fully developed unsteady regime \cite{DeBlois2023, Mokhtari2024}. 

\subsubsection{Steady-State Stress Structures}
\begin{figure}[!ht]
    \centering
    \includegraphics[width=\columnwidth]{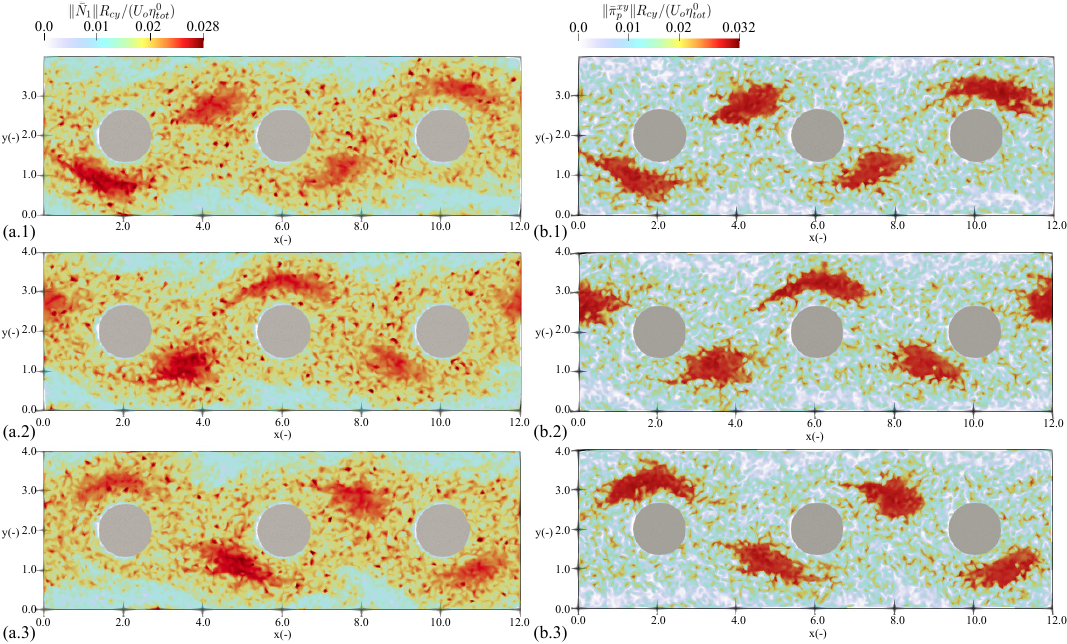}
    \caption{Steady-state coherent stress structures and elastically driven mixing. Dimensionless (a) $\bar{N}_1$ and (b) $\|\bar{\pi}^{xy}_p\|$ fields. Sequential snapshots at intervals of $\Delta t \approx 0.4/\Omega_{\text{peak}}$ ($t \approx \lambda + 0.4/\Omega_{\text{peak}}, \lambda + 0.8/\Omega_{\text{peak}}, \lambda + 1.2/\Omega_{\text{peak}}$) illustrate the traveling-wave propagation of the stress fields and sustained stretching--relaxation cycles, where $\Omega_{\text{peak}}$ is defined in Fig.~\ref{fig:psd_multi}(b) for the $3\text{L} \times \text{H}$ array.}
    \label{fig1-SS}
\end{figure}

In the developed oscillatory regime (Fig.~\ref{fig1-SS}), the stress fields organize into a periodic traveling-wave pattern formed by concentrated clusters of $\bar{N}_1$ and $\|\bar{\pi}^{xy}_p\|$. These stress lobes oscillate in phase with the transverse velocity $v_y$ and the undulating streamlines (Fig.~\ref{fig:multi_obstacle_flow}(a)). In multi-obstacle arrays, the interaction between adjacent wakes broadens the frequency spectrum, as detailed in Appendices~\ref{sec:results_1PAC_PSD} and \ref{sec:scalabilityEI}.

The fluctuations shown in Fig.~\ref{fig1-SS} enhance advective mixing by inducing a transverse ``squeezing'' effect on the streamlines (Fig.~\ref{fig:multi_obstacle_flow}(a)). Although $\bar{N}_1$ and $\|\bar{\pi}^{xy}_p\|$ peaks coincide spatially, they exhibit distinct morphologies: the first normal stress ($\bar{N}_1$, Fig.~\ref{fig1-SS}(a)) forms broad, undulating ``polymeric tension sheets'' that flap transversely as they propagate, reflecting intense chain elongation near stagnation points. By contrast, the shear stress ($\|\bar{\pi}^{xy}_p\|$, Fig.~\ref{fig1-SS}(b)) concentrates into sharper filaments wrapping around the cylinder perimeters, corresponding to the birefringent strands reported in experiments and continuum simulations \cite{Mokhtari2024}. 

Unlike classical continuum formulations \cite{Mokhtari2022, Mokhtari2023} that require artificial stress diffusion or Peterlin-type closure approximations to regularize high-$\text{Wi}$ singularities, the LHMM captures these sticky flow barriers directly from the bottom-up coupling. Local SPH velocity gradients deform the discrete FENE chains inside each DPD micro-domain, and the resulting ensemble-averaged extra-stresses feed back into the macroscopic momentum balance \cite{Patino-Narino2025a, Moreno2026a}. Localized flow resistance thus emerges directly from history-dependent chain elongation at trailing stagnation points, which deflects the fluid along preferential transport channels \cite{FoggiRota2026} and sustains continuous cycles of stretching and relaxation across the interstitial throats.

\subsection{Onset of Instability and Bifurcation Analysis: Effect of the Weissenberg Number (Wi)}\label{sec:bifurcation_analysis}
To map the stability threshold and characterize the nonlinear transition across the multiscale framework, we varied the driving body force over the range $8.5 \times 10^{-4} \le F \le 3.0 \times 10^{-2}$ at a fixed polymer viscosity ratio $\beta_p = 0.97$ and spatial resolution $R_{\text{cy}}/\Delta x = 7.0$ in the $3\text{L} \times \text{H}$ array. This sweep spans $2.2 \le \text{Wi} \le 130$ and $8.3 \times 10^{-4} \le \text{Re} \le 0.05$.

\begin{figure}[!ht]   
    \centering
    \includegraphics[width=\columnwidth]{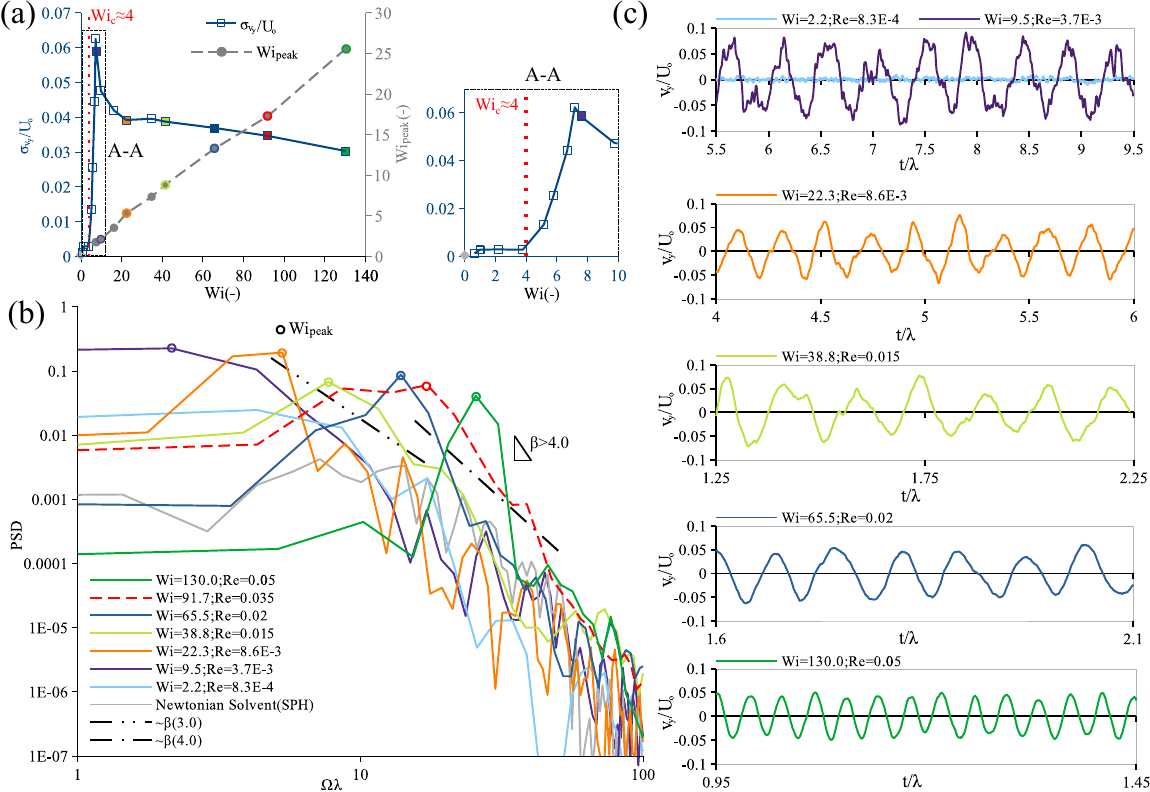} 
    \caption{Bifurcation analysis and spectral onset of the purely elastic instability in a $3\text{L} \times \text{H}$ array. 
    (a) Bifurcation diagram showing normalized transverse velocity fluctuations $\sigma_{v_y}/U_0$ (left axis) and dominant frequency parameter $\text{Wi}_{\text{peak}}$ (right axis) versus $\text{Wi}$. Inset A-A highlights the onset of self-sustained oscillations at $\text{Wi}_c \approx 4.0$, marking the transition from a steady symmetric state to an unsteady oscillatory regime.
    (b) Power Spectral Density (PSD) of transverse velocity fluctuations ($2.2 \le \text{Wi} \le 130$). Dashed-dotted lines indicate reference power-law slopes ($\beta = 3$ and $4$), tracing the transition from a featureless state ($\text{Wi}=2.2$) to broadband, multi-frequency fluctuations consistent with elastic-turbulence-like scaling. At extreme elasticity ($\text{Wi} = 130$), the spectrum exhibits a steep drop-off ($\beta > 4$) due to harmonic regularization. The reference case ($\text{Wi}=91.7$, $\text{Re}=0.035$) is highlighted by a red dashed line.
    (c) Temporal signals of normalized transverse velocity $v_y/U_0$ across representative cases ($\text{Wi} \in \{2.2, 9.5, 22.3, 38.8, 65.5, 130\}$), illustrating waveform distortion near onset and progressive harmonic regularization at high $\text{Wi}$.}
    \label{fig:bifurcation}
\end{figure}

Figure~\ref{fig:bifurcation}(a) presents the global bifurcation diagram, tracking the normalized standard deviation of the transverse velocity fluctuations ($\sigma_{v_y}/U_0$, left axis) and the dimensionless dominant frequency ($\text{Wi}_{\text{peak}} = \Omega_{\text{peak}}\lambda$, right axis) as functions of $\text{Wi}$ in the statistically stationary state. For $\text{Wi} \le 4.0$, the flow remains in a stable, spatially symmetric state closely matching the Newtonian solvent baseline ($\sigma_{v_y}/U_0 \approx 0$) \cite{Patino-Narino2025a, Datta2022}.

As $\text{Wi}$ exceeds the critical threshold $\text{Wi}_c \approx 4.0$ (inset A-A of Fig.~\ref{fig:bifurcation}(a)), the flow loses steady-state symmetry and transitions to self-sustained oscillations \cite{Mokhtari2024, Beneitez2024} as elastic stress generation in the stagnation zones overcomes viscous dissipation \cite{Berti2008, Zhu2024}. Above onset, the fluctuation amplitude $\sigma_{v_y}/U_0$ rises rapidly, peaks within $7.0 \le \text{Wi} \le 9.5$, and gradually attenuates at higher elasticity ($\text{Wi} \to 130$) as strong streamwise advection aligns the elastic tension forces along the main flow direction. Meanwhile, the dominant frequency parameter $\text{Wi}_{\text{peak}}$ grows almost linearly from $2.2$ to $25.6$ (Fig.~\ref{fig:bifurcation}(a)), confirming that the transverse whip-like oscillations are governed by advective transport and cyclic chain stretching across the wakes \cite{Nichols2025a, Haward2021c}.

The corresponding Power Spectral Density (PSD) profiles in Fig.~\ref{fig:bifurcation}(b) and velocity traces in Fig.~\ref{fig:bifurcation}(c) capture the spectral evolution across $\text{Wi}$. Above onset ($9.5 \le \text{Wi} \le 91.7$), multi-frequency wake interactions produce distorted, non-sinusoidal waveforms and a broadband high-frequency spectral decay with apparent power-law slopes $\beta \approx 3\text{--}4$ over the resolved inertialess range, consistent with temporal scaling exponents reported for elastic turbulence \cite{Groisman2004a, Berti2008, FoggiRota2026}. As detailed in Appendices~\ref{sec:results_1PAC_PSD} and \ref{sec:scalabilityEI}, this broadband spectral decay requires extended multi-obstacle arrays ($n \ge 3$) to allow cumulative wake interference across cells. At extreme elasticity ($\text{Wi} = 130$), however, intense mean-flow advection suppresses high-frequency fluctuations: the waveform regularizes into a nearly harmonic oscillation dominated by a sharp fundamental peak at $\text{Wi}_{\text{peak}} = 25.6$ with a steep high-frequency drop-off ($\beta > 4$), restoring the multi-obstacle array to a quasi-periodic limit cycle.

\begin{figure}[!ht]
    \centering
    \includegraphics[width=\textwidth]{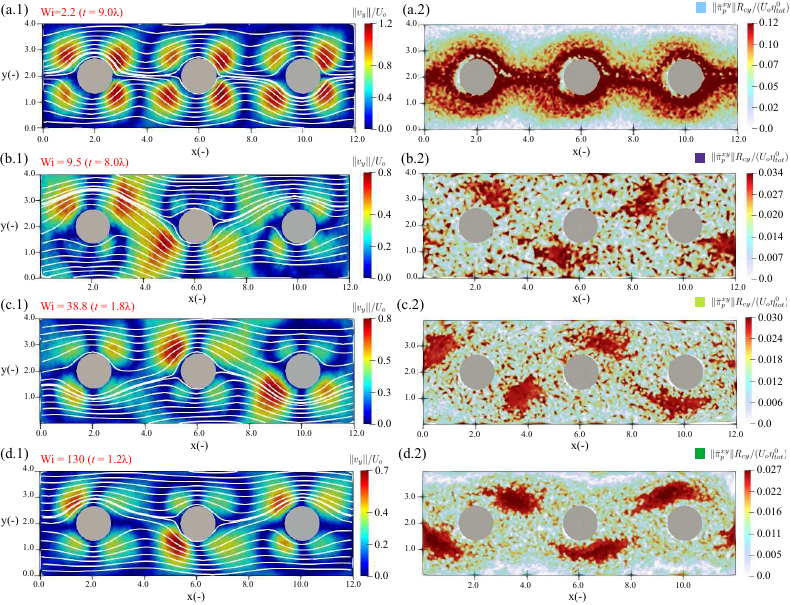}
    \caption{Steady-state flow topology and polymeric shear stress distributions across distinct stability regimes. 
    (a) Stable case $\text{Wi}=2.2$ ($\text{Re}=8.3 \times 10^{-4}$): (a.1) $|v_y|/U_0$ with streamlines, and (a.2) $\|\bar{\pi}_p^{xy}\|$ field at $t=9.0\lambda$.
    (b) $\text{Wi}=9.5$ ($\text{Re}=3.7 \times 10^{-3}$): (b.1) $|v_y|/U_0$ with streamlines, and (b.2) $\|\bar{\pi}_p^{xy}\|$ field at $t=8.0\lambda$.
    (c) $\text{Wi}=38.8$ ($\text{Re}=0.015$): (c.1) $|v_y|/U_0$ with streamlines, and (c.2) $\|\bar{\pi}_p^{xy}\|$ field at $t=1.8\lambda$.
    (d) $\text{Wi}=130$ ($\text{Re}=0.05$): (d.1) $|v_y|/U_0$ with streamlines, and (d.2) $\|\bar{\pi}_p^{xy}\|$ field at $t=1.2\lambda$.}
    \label{fig:BodyF_profiles}
\end{figure}

Figure~\ref{fig:BodyF_profiles} links these spectral regimes to the spatial distributions of $|v_y|/U_0$ (with streamlines) and the polymeric shear stress $\|\bar{\pi}_p^{xy}\|$. In the subcritical regime ($\text{Wi} = 2.2$, Fig.~\ref{fig:BodyF_profiles}(a)), both fields remain symmetric with unperturbed streamlines and continuous stress shells around the obstacles. At intermediate post-critical values ($\text{Wi} = 9.5$ and $38.8$, Figs.~\ref{fig:BodyF_profiles}(b,c)), the streamlines undulate and the polymeric stress fragments into localized birefringent strands wrapping the cylinders \cite{Mokhtari2024}. At $\text{Wi} = 130$ (Fig.~\ref{fig:BodyF_profiles}(d)), intense streamwise advection broadens and partially elongates these stress clusters along the main flow channel, consistent with the amplitude saturation of $\sigma_{v_y}/U_0$ in Fig.~\ref{fig:bifurcation}(a).

\subsection{Stability Analysis and Flow Dynamics: Effect of the Polymer Viscosity Ratio (\texorpdfstring{$\beta_p$}{beta\_p})}\label{app:dilute_limit}
We next examine the influence of the polymer viscosity ratio ($\beta_p$) on flow stability by varying $\beta_p$ from $0.77$ to $0.99$ under a constant driving body force ($F = 0.02$). This sweep isolates how the relative polymeric contribution to the total zero-shear viscosity controls flow destabilization while maintaining local $\text{Wi}$ values well above $\text{Wi}_c \approx 4.0$.

\begin{figure}[!ht]
    \centering
    \includegraphics[width=\textwidth]{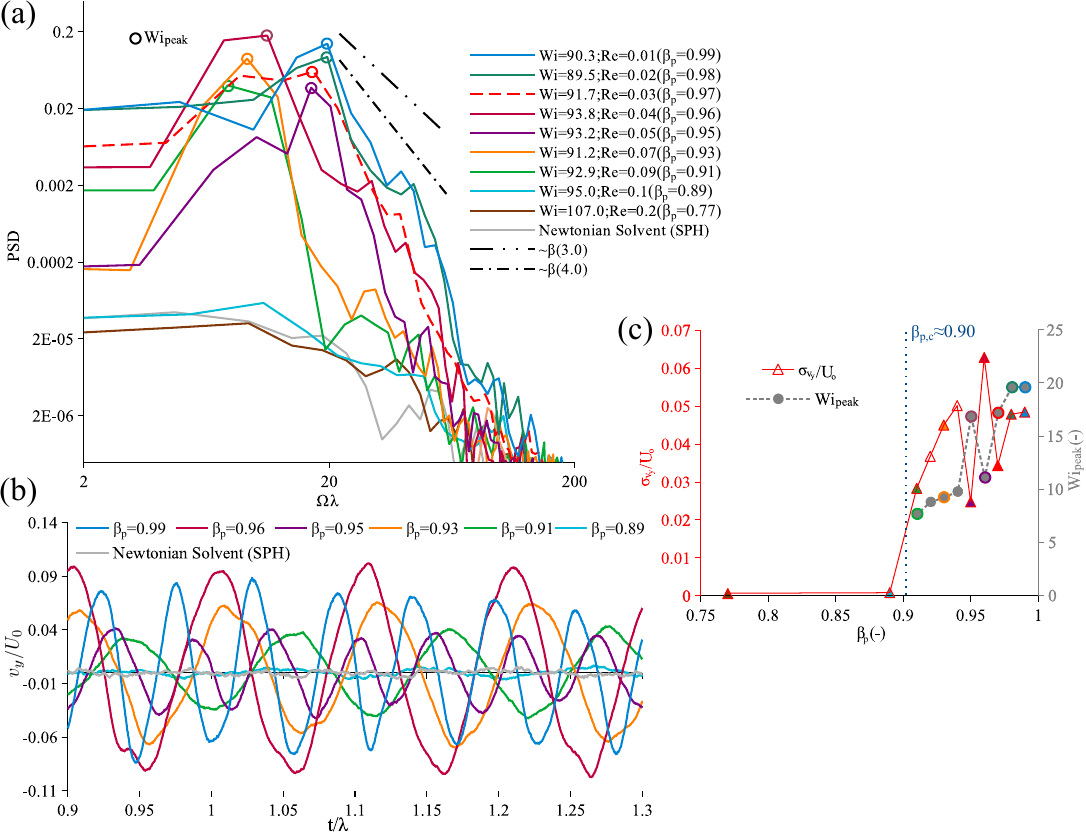}
    \caption{Temporal and spectral characterization of the concentration-driven stability threshold across viscosity ratio ($\beta_p$) variations. 
    (a) PSD of transverse velocity fluctuations for $0.77 \le \beta_p \le 0.99$; the emergence of a dominant harmonic peak at $\beta_p \ge 0.90$ identifies the critical concentration boundary ($\beta_{p,c} \approx 0.90$). Dashed-dotted lines representing $\beta = 3$ and $\beta = 4$ decays indicate reference power-law slopes typically reported for elastic turbulence.
    (b) Temporal evolution of normalized transverse velocity $v_y/U_0$ across six cases ($0.89 \le \beta_p \le 0.99$) during $0.9\lambda \le t \le 1.3\lambda$, illustrating signal distortion and amplitude modulation with increasing $\beta_p$. 
    (c) Stability diagram displaying the normalized transverse velocity fluctuation intensity $\sigma_{v_y}/U_0$ (left axis) and dominant frequency parameter $\text{Wi}_{\text{peak}}$ (right axis) against $\beta_p$ at constant forcing ($F = 0.02$). The flow remains stable ($\sigma_{v_y}/U_0 \approx 0$) for $\beta_p < 0.90$.}
    \label{fig:app_temporal}
\end{figure}

Figure~\ref{fig:app_temporal}(a) displays the PSD of transverse velocity fluctuations across $\beta_p$. Below the critical boundary ($\beta_p < \beta_{p,c} \approx 0.90$), the spectra remain flat and collapse onto the Newtonian baseline, showing that solvent viscous dissipation suppresses elastic energy accumulation at lower polymer contributions. Crossing $\beta_{p,c} \approx 0.90$ triggers a sharp transition marked by the emergence of distinct spectral peaks. For intermediate ratios ($0.90 \le \beta_p < 0.96$), the response is governed by limit-cycle oscillations, whereas for $\beta_p \ge 0.96$ these discrete peaks broaden into a continuous spectrum exhibiting apparent power-law decay slopes $\beta \approx 3\text{--}4$. This trend also rules out inertial effects: although reducing $\beta_p$ raises $\text{Re}$ from $0.01$ ($\beta_p = 0.99$) to $0.2$ ($\beta_p = 0.77$) and shear thinning further lowers the local viscosity, the flow stabilizes precisely at the highest $\text{Re}$.

Figure~\ref{fig:app_temporal}(b) shows the corresponding time series of $v_y/U_0$ during $0.9\lambda \le t \le 1.3\lambda$. Below threshold ($\beta_p = 0.89$), fluctuations decay to zero; above threshold ($\beta_p \ge 0.90$), the signal develops nonlinear distortions and large-amplitude oscillations. The bifurcation diagram in Fig.~\ref{fig:app_temporal}(c) quantifies this transition: $\sigma_{v_y}/U_0$ remains at the noise floor for $\beta_p < 0.90$, grows monotonically across $0.91 \le \beta_p \le 0.94$, exhibits a local mode shift at $\beta_p = 0.95$ (where $\sigma_{v_y}/U_0$ dips while $\text{Wi}_{\text{peak}}$ jumps), reaches its global maximum at $\beta_p = 0.96$, and plateaus for $\beta_p \ge 0.97$ as stress reorganization balances elastic energy injection \cite{Patino-Narino2025a}.

\begin{figure}[!ht]
    \centering
    \includegraphics[width=\textwidth]{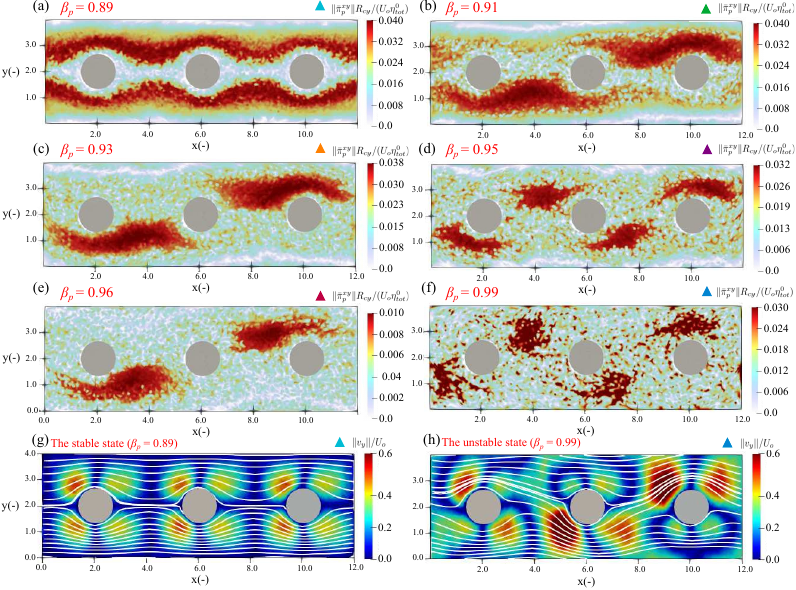}
    \caption{Spatial evolution of normalized polymeric shear stress $\|\bar{\pi}_p^{xy}\| R_{\text{cy}} / (U_0 \eta_{\text{tot}}^0)$ and steady-state flow topology at $t = 1.4\lambda$. 
    (a--f) Spatial stress distribution $\|\bar{\pi}_p^{xy}\|$ for $\beta_p = \{0.89, 0.91, 0.93, 0.95, 0.96, 0.99\}$, respectively. 
    (g--h) Transverse velocity contours $|v_y|/U_0$ overlaid with streamlines for (g) the stable state ($\beta_p = 0.89$), showing symmetric streamlines, and (h) the unstable state ($\beta_p = 0.99$), showing broken symmetry.}
    \label{fig:app_stress_evol}
\end{figure}

The spatial stress fields $\|\bar{\pi}_p^{xy}\|$ in Figs.~\ref{fig:app_stress_evol}(a--f) and velocity contours in Figs.~\ref{fig:app_stress_evol}(g,h) illustrate the structural origin of this threshold. At subcritical concentration ($\beta_p = 0.89$, Figs.~\ref{fig:app_stress_evol}(a,g)), the stress field forms symmetric arrowhead-like structures anchored at the trailing stagnation points with unperturbed rectilinear streamlines, consistent with steady Exact Coherent Structures \cite{Zhu2025, Lellep2023, Morozov2022}. Above threshold, the stress bands fragment into localized clusters at intermediate $\beta_p$ ($0.91\text{--}0.93$, Figs.~\ref{fig:app_stress_evol}(b,c)) and organize into elongated shear bands spanning the constriction channels at $\beta_p = 0.96$ (Fig.~\ref{fig:app_stress_evol}(e)), driving the peak oscillation amplitude in Fig.~\ref{fig:app_temporal}(b)--(c) and the strongly undulating streamlines at $\beta_p = 0.99$ (Fig.~\ref{fig:app_stress_evol}(h)).

\subsection{Effect of Array Porosity} \label{sec:results_porosity}
We conclude the parametric analysis by evaluating the effect of array porosity ($\varepsilon = 1 - \pi R_{\mathrm{cy}}^2 / (\text{LH})$) in the $3\text{L} \times \text{H}$ domain ($\text{H}/\Delta x = 50.0$) across five unstable configurations spanning three porosities: $\varepsilon = 0.91$, obtained by shortening the longitudinal pitch ($\text{L} = 2.8$) at fixed obstacle radius ($R_{\mathrm{cy}} = 0.56$, blockage ratio $2R_{\mathrm{cy}}/\text{H} = 0.28$), alongside $\varepsilon = 0.80$ ($R_{\mathrm{cy}} = 1.0$, $2R_{\mathrm{cy}}/\text{H} = 0.50$) and $\varepsilon = 0.69$ ($R_{\mathrm{cy}} = 1.25$, $2R_{\mathrm{cy}}/\text{H} = 0.625$) at fixed cell spacing ($\text{L} = \text{H} = 4.0$). Because modifying either the pitch or the transverse blockage under fixed driving forces simultaneously alters the interstitial velocity and local shear rates, geometric confinement and $\text{Wi}$ co-vary across these cases ($6.0 \le \text{Wi} \le 56.0$). We therefore examine how their combined interplay alters the spectral and spatial organization of the unstable flow.

\begin{figure}[!ht]
    \centering
    \includegraphics[scale=0.95]{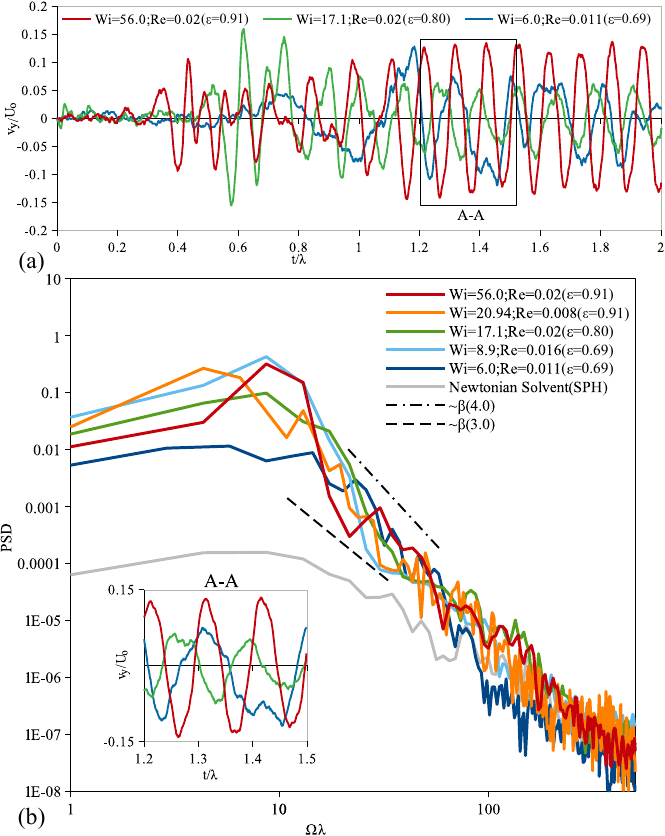}
    \caption{Dynamic response and spectral broadening across different porosities in $3\text{L} \times \text{H}$ arrays. 
    (a) Temporal evolution of normalized transverse velocity $v_y/U_0$ for $\varepsilon = 0.91$, $0.80$, and $0.69$, with inset A-A showing waveform distortion for $t > 1.2\lambda$. 
    (b) PSD for five configurations illustrating broadband spectral decay with apparent slopes ($\beta \approx 3\text{--}4$) above the dominant frequency.}
    \label{fig:porosity_cases} 
\end{figure}

As shown in Fig.~\ref{fig:porosity_cases}(a)--(b), stronger transverse confinement promotes broadband disorder even at markedly lower $\text{Wi}$. In tightly blocked arrays ($\varepsilon = 0.69$, $\text{Wi}=6.0$ and $8.9$), narrow pore constrictions disrupt coherent wake oscillations, producing anharmonic velocity fluctuations (inset A-A) and a broad spectral decay ($\beta \approx 3\text{--}4$) without a sharp harmonic peak. By contrast, in the higher-porosity, lower-blockage lattice ($\varepsilon = 0.91$, $\text{Wi}=20.9$ and $56.0$), wider transverse clearance allows elastic structures to advect past successive obstacles with less geometric obstruction, preserving prominent harmonic peaks before decaying at higher frequencies. Because the tightest geometry produces broader, more anharmonic fluctuations at much lower $\text{Wi}$, transverse constriction plays a stronger role than elasticity alone in disrupting temporal coherence.

\begin{figure}[!ht]
    \centering
    \includegraphics[width=\columnwidth]{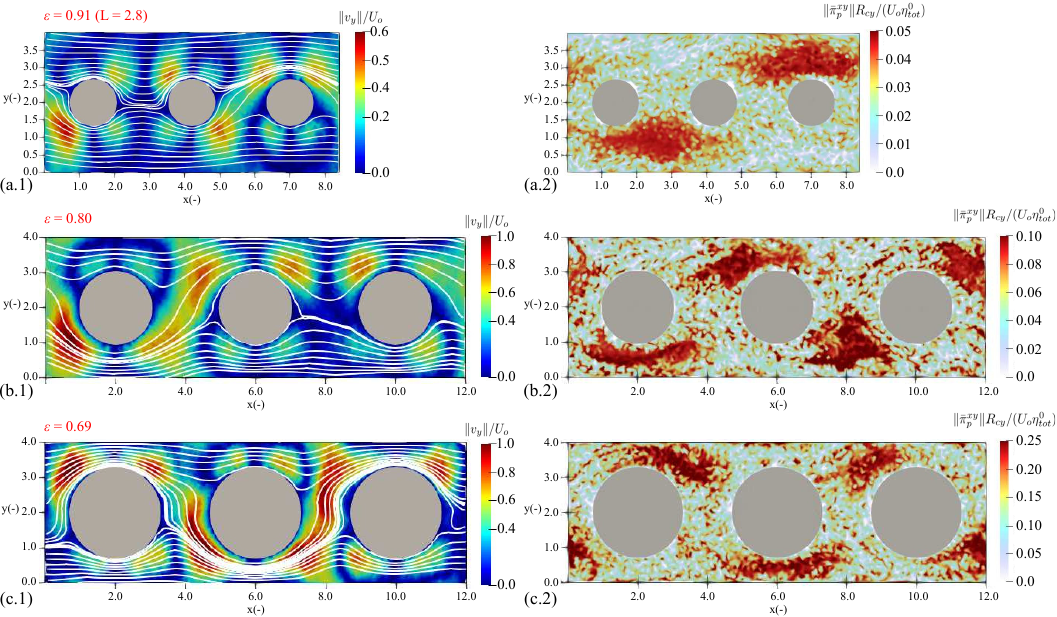}
    \caption{Comparative flow topology and polymeric shear stress distributions across porosities in $3\text{L} \times \text{H}$ arrays. 
    (a) $\varepsilon = 0.91$ ($\text{L} = 2.8$), (b) $\varepsilon = 0.80$, and (c) $\varepsilon = 0.69$. Panels (.1) show streamlines colored by $|v_y|/U_0$ for $t > \lambda$; panels (.2) display corresponding polymeric shear stress fields $\|\bar{\pi}^{xy}_p\| R_{\text{cy}} / (U_0 \eta_{\text{tot}}^0)$, comparing localized stress pinning in low-porosity media with strand elongation in high-porosity lattices.}
    \label{fig:porosity_visual}
\end{figure}

Figure~\ref{fig:porosity_visual} illustrates the spatial mechanism behind both regimes. In tightly confined media ($\varepsilon = 0.80$ and $0.69$, Figs.~\ref{fig:porosity_visual}(b,c)), narrow interstitial throats restrict wake advection, forcing high-stress pockets to remain pinned within the constriction gaps \cite{Dubief2022}. At higher porosity ($\varepsilon = 0.91$, Fig.~\ref{fig:porosity_visual}(a)), the wider transverse clearance enables the birefringent strands to elongate and advect across successive cylinders, sustaining large-scale transverse whip-like oscillations \cite{Haward2021c, Mokhtari2024}. Geometric confinement thus dictates whether the microscopic stress barriers remain trapped inside pore throats or evolve into mobile, advected filaments.

\subsection{Microstructural Evolution and Conformational Statistics at the Onset of EI} \label{sec:conformational_metrics}
In this section, we examine the effect of microstructural polymer changes on the onset of EI by inspecting a moderate-forcing regime ($F = 0.01$, corresponding to $\text{Wi} = 38.8$ and $\text{Re} = 0.015$), which provides high temporal resolution across the transition. To characterize the microstructural response and conformational dynamics of the FENE polymer chains under flow, we extract the gyration tensor $\mathbf{R}'_g$ directly from the bead-spring configurations within the DPD micro-domains ($\Omega'_{\mathrm{core}}$). To prevent artificial chain truncation across the core--buffer interface, molecular chunks are first evaluated across the entire micro-domain and subsequently mapped onto the core volume. For an individual polymer chain composed of $N_b = 16$ beads with coordinates $\mathbf{r}'_i$, the instantaneous polymer center of mass $\mathbf{r}'_{\mathrm{cm}}$ and the components of the gyration tensor are defined as:
\begin{equation} \label{eq:gyration_tensor}
    \mathbf{r}'_{\mathrm{cm}} = \frac{1}{N_b} \sum_{i=1}^{N_b} \mathbf{r}'_i, \qquad
    R'_{g,\alpha\beta} = \frac{1}{N_b} \sum_{i=1}^{N_b} \left(r'_{i,\alpha} - r'_{\mathrm{cm},\alpha}\right)\left(r'_{i,\beta} - r'_{\mathrm{cm},\beta}\right),
\end{equation}
where $\alpha, \beta \in \{x, y\}$. 

In the macroscopic flow plane ($x\text{--}y$), the collective conformational state associated with each Lagrangian SPH fluid particle is obtained via ensemble averaging over the micro-domain, yielding the planar 2D gyration tensor:
\begin{equation} \label{eq:ensemble_rg_tensor}
    \mathbf{R}_g = \langle \mathbf{R}'_g \rangle = 
    \begin{bmatrix}
        R_{g,xx} & R_{g,xy} \\
        R_{g,xy} & R_{g,yy}
    \end{bmatrix}.
\end{equation}
Here, the diagonal components $R_{g,xx}$ and $R_{g,yy}$ quantify normal molecular stretching along the streamwise (flow) and cross-stream (gradient) directions, respectively, whereas the off-diagonal term $R_{g,xy}$ captures the conformational shear deformation and collective chain tilt induced by velocity gradients. 

From the rotational trace invariant of $\mathbf{R}_g$, we determine the scalar radius of gyration:
\begin{equation} \label{eq:rg_scalar_trace}
    R_{g} = \sqrt{\mathrm{Tr}(\mathbf{R}_g)} = \sqrt{R_{g,xx} + R_{g,yy}},
\end{equation}
which provides an orientation-independent measure of overall chain extension. In addition to the ensemble mean, the sub-particle conformational spread within each SPH macro-particle is tracked via the intrinsic population variance:
\begin{equation} \label{eq:micro_variance}
    \sigma_{\mathrm{micro}}^2 = \langle R'^2_{g} \rangle - \langle R'_{g} \rangle^2,
\end{equation}
which directly quantifies microscopic conformational heterogeneity (the coexistence of stretched and tumbling molecular states) independently of the macroscopic mean.

\begin{figure}[!ht]
    \centering
    \includegraphics[width=\columnwidth]{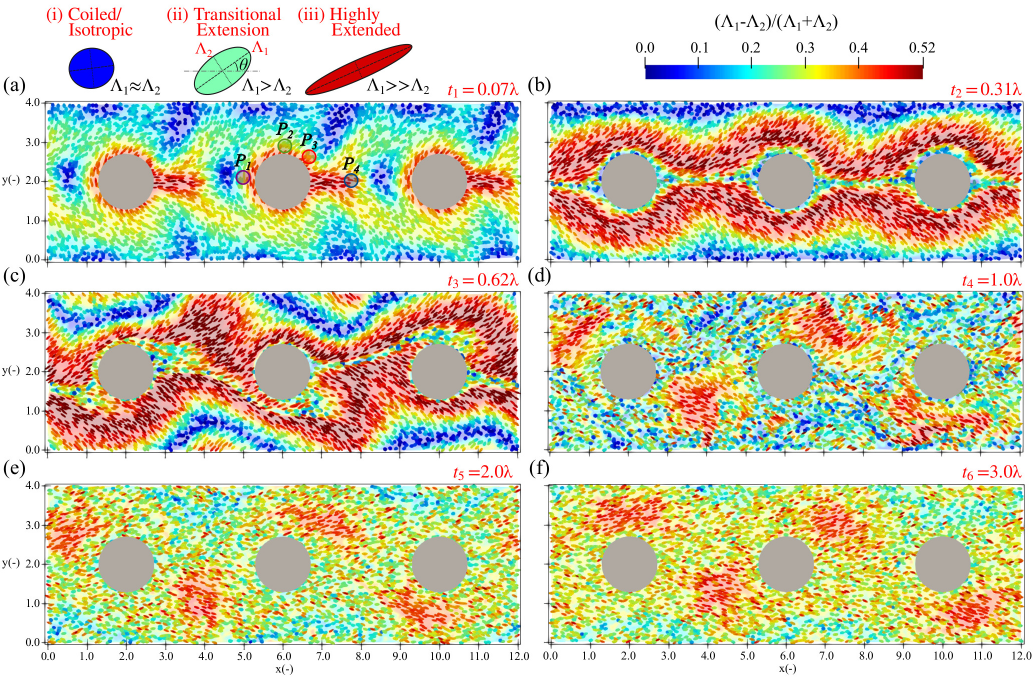}
    \caption{Spatiotemporal evolution of the polymer conformation field across the three-cylinder array for $\text{Wi} = 38.8$ ($\text{Re} = 0.015, F = 0.01$) at representative dimensionless times: (a) $t_1 = 0.07\lambda$ (indicating the four Eulerian monitoring probes: $P_1(5.1, 1.95)$, $P_2(6.3, 2.80)$, $P_3(6.7, 2.45)$, and $P_4(8.0, 2.05)$), (b) $t_2 = 0.31\lambda$, (c) $t_3 = 0.62\lambda$, (d) $t_4 = 1.0\lambda$, (e) $t_5 = 2.0\lambda$, and (f) $t_6 = 3.0\lambda$. Each SPH macro-particle is rendered as a 2D tensor glyph (ellipsoid) whose semi-axes and tilt angle reflect the eigenvalues ($\Lambda_1, \Lambda_2$) and eigenvectors ($\theta$) of $\mathbf{R}_g$, colored by the dimensionless planar anisotropy index $(\Lambda_1 - \Lambda_2)/(\Lambda_1 + \Lambda_2)$. The top schematic illustrates the glyph morphology across (i) coiled/isotropic ($\Lambda_1 \approx \Lambda_2$), (ii) transitional extension ($\Lambda_1 > \Lambda_2$), and (iii) highly extended ($\Lambda_1 \gg \Lambda_2$) states. In all panels, subsampled tensor glyphs are overlaid on a semitransparent $(\Lambda_1 - \Lambda_2)/(\Lambda_1 + \Lambda_2)$ surface to enhance the visual contrast of the coherent structures.}
    \label{fig:rg_evolution_3cyl}
\end{figure}

In the spatial visualizations (Figs.~\ref{fig:rg_evolution_3cyl} and \ref{fig:rg_details_p1}, and \hyperref[SM]{supplementary animations} \texttt{Rg.mp4} and \texttt{Rg-Zoom.mp4}), the planar tensor $\mathbf{R}_g$ is diagonalized locally at each SPH macro-particle position ($\mathbf{R}_g \xrightarrow{\theta} \mathrm{diag}(\Lambda_1, \Lambda_2)$) and rendered using 2D tensor glyphs (ellipsoids) in ParaView~\cite{Ahrens2005}. Each glyph encodes three local observables, as schematically illustrated at the top of Fig.~\ref{fig:rg_evolution_3cyl}: (i) the major and minor semi-axes scale with the principal eigenvalues ($\Lambda_1, \Lambda_2$) of $\mathbf{R}_g$, distinguishing coiled/isotropic states ($\Lambda_1 \approx \Lambda_2$, nearly circular glyphs) and transitional extension ($\Lambda_1 > \Lambda_2$, moderately eccentric ellipsoids) from highly extended states ($\Lambda_1 \gg \Lambda_2$, strongly elongated ellipsoids); (ii) the major-axis orientation aligns with the principal eigenvector angle $\theta = \frac{1}{2}\arctan\left[2R_{g,xy} / (R_{g,xx} - R_{g,yy})\right]$; and (iii) the surface color maps the dimensionless planar conformational anisotropy index $(\Lambda_1 - \Lambda_2)/(\Lambda_1 + \Lambda_2)$ in Fig.~\ref{fig:rg_evolution_3cyl} (spanning from $0$ for isotropic coiled states to $1$ for fully extended chains), whereas the close-up view in Fig.~\ref{fig:rg_details_p1} maps the normalized scalar radius of gyration $R_{g}/\langle R_g \rangle$, where $\langle R_g \rangle$ is the instantaneous spatial mean over the entire fluid domain.

Figure~\ref{fig:rg_evolution_3cyl} illustrates the macroscopic sequence of the conformational transition across the channel from $t_1 = 0.07\lambda$ to $t_6 = 3.0\lambda$, corroborating the stress evolution observed at higher forcing (Figs.~\ref{fig:transient_evolution} and \ref{fig1-SS}). During early start-up at $t_1 = 0.07\lambda$ (panel (a)), elongated red ellipsoids concentrate along the cylinder boundary layers and form straight, symmetric birefringent strands connecting the trailing and leading stagnation points of consecutive obstacles, while macro-particles in the outer channel remain blue and nearly circular. As the quasi-steady state develops at $t_2 = 0.31\lambda$ (panel (b)), these high-anisotropy bands split and extend outward over the upper and lower shoulders of the cylinders into the constriction gaps, leaving a narrow, weakly anisotropic (blue/green) strip along the inter-cylinder centerline. These high-anisotropy filaments form the microstructural backbone of the macroscopic shear-stress strands ($\|\bar{\pi}_p^{xy}\|$) and polymeric tension sheets ($\bar{N}_1$) identified in Fig.~\ref{fig:transient_evolution}(a). At $t_3 = 0.62\lambda$ (panel (c)), elastic saturation along the curved boundaries breaks the wake symmetry, causing the birefringent strand to buckle and oscillate transversely. By $t_4 = 1.0\lambda$ (panel (d)), the shedding and fragmentation of these unrelaxed wakes initiate the cooperative mixing effect across the pore space.

As the system settles into the statistically stationary regime at $t_5 = 2.0\lambda$ and $t_6 = 3.0\lambda$ (panels (e)--(f)), the spatial distribution of the tensor glyphs reproduces the steady-state stress morphology of Fig.~\ref{fig1-SS}. Instead of relaxing into a uniform disordered field, the conformational field self-organizes into a periodic traveling-wave pattern: alternating, oblique clusters of stretched and tilted orange/red ellipsoids span the upper and lower inter-obstacle gaps, providing the direct chain-level realization of the flapping tension sheets ($\bar{N}_1$) and sticky flow barriers ($\|\bar{\pi}_p^{xy}\|$). These high-extension barriers are surrounded by cyan-to-yellow, less deformed glyphs traversing the constriction throats, visually confirming the elastically driven squeezing mechanism that routes the fluid along oscillating transport pathways.

\begin{figure}[!ht]
    \centering
    \includegraphics[scale=0.8]{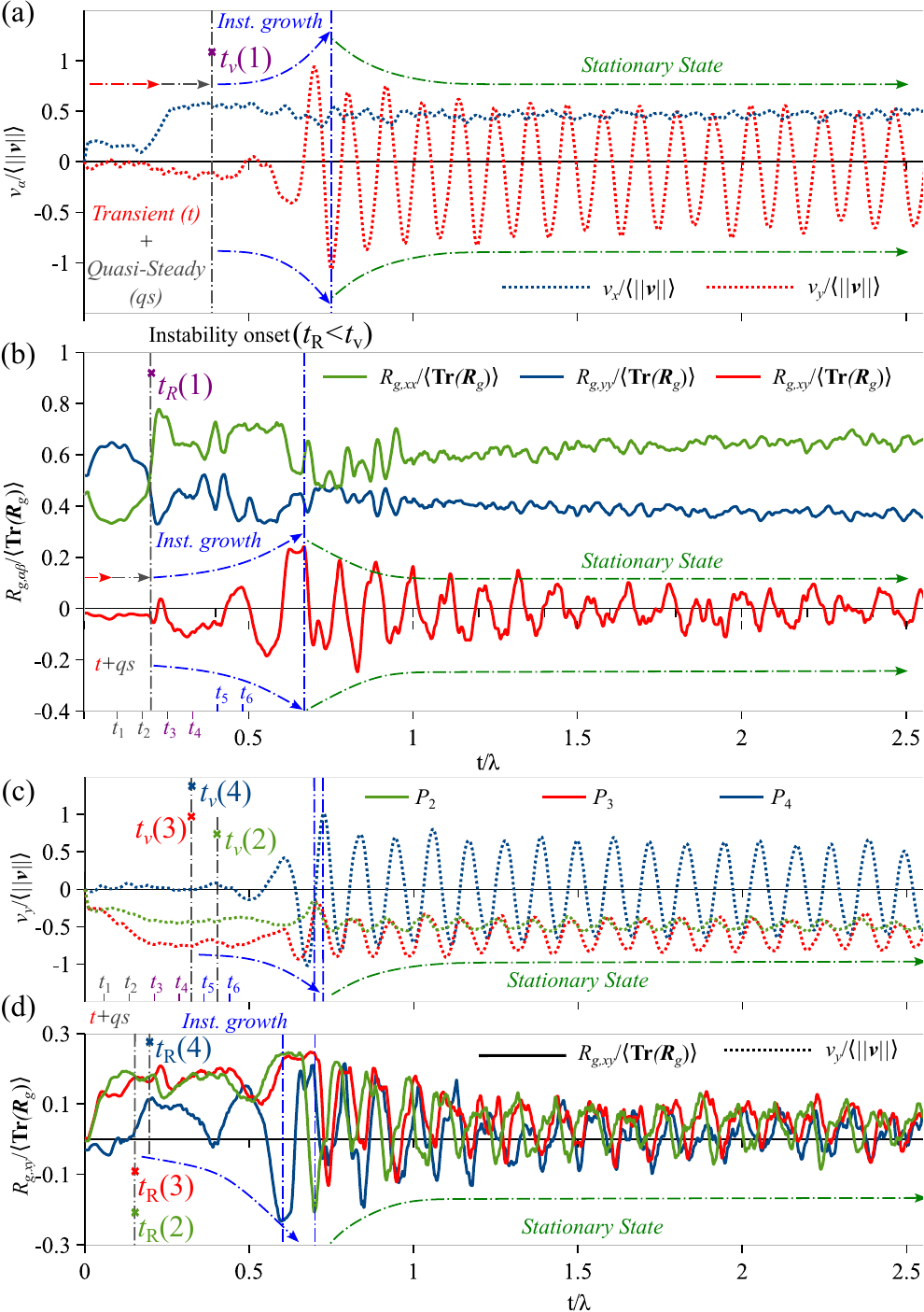}
    \caption{Coupled temporal evolution of normalized flow kinematics and polymer conformation metrics across the four control probes ($P_1\text{--}P_4$) for $\text{Wi} = 38.8$ ($\text{Re} = 0.015$). The top two panels detail the response at the upstream probe $P_1(5.1, 1.95)$: normalized velocity components ($v_x/\langle\|\mathbf{v}\|\rangle$ and $v_y/\langle\|\mathbf{v}\|\rangle$ in panel a) and normalized gyration tensor components ($R_{g,xx}, R_{g,yy}, R_{g,xy}$ scaled by $\langle\mathrm{Tr}(\mathbf{R}_g)\rangle$ in panel b). The bottom two panels compare the transverse velocity fluctuations ($v_y / \langle \|\mathbf{v}\| \rangle$, panel c) and the normalized conformational shear component ($R_{g,xy} / \langle \mathrm{Tr}(\mathbf{R}_g) \rangle$, panel d) across downstream probes $P_2(6.3, 2.80)$, $P_3(6.7, 2.45)$, and $P_4(8.0, 2.05)$. Tick marks $t_1\text{--}t_6$ along the horizontal axes of panels (b) and (c) indicate the sampling instants analyzed in Fig.~\ref{fig:rg_statistics}. Vertical dash-dotted lines mark the conformational onset times $t_R(i)$ versus the transverse kinematic onset times $t_v(i)$, demonstrating that microstructural tensor reorganization precedes macroscopic lateral flow oscillations ($t_R(i) < t_v(i)$).}
    \label{fig:rg_vs_vel}
\end{figure}

Figure~\ref{fig:rg_vs_vel} provides a time-resolved comparison between local fluid kinematics and internal microstructural dynamics at the four Eulerian control probes ($P_1\text{--}P_4$, indicated in Fig.~\ref{fig:rg_evolution_3cyl}(a)), resolving four dynamical stages: an initial start-up transient ($t$), a quasi-steady plateau ($qs$), an instability growth phase, and a statistically stationary fluctuating regime. Here, the onset times $t_R(i)$ and $t_v(i)$ are defined as the instants at which the local gyration tensor components and the transverse velocity $v_y$, respectively, exhibit a sustained departure from their quasi-steady ($qs$) symmetric baselines (exceeding the initial baseline noise envelope). Note that while $v_x$ at $P_1$ begins adjusting alongside the conformational shift near $t_R(1) \approx 0.22\lambda$, $t_v(i)$ specifically tracks the onset of lateral symmetry breaking ($v_y \neq 0$).

A key result of Fig.~\ref{fig:rg_vs_vel} is a consistent temporal offset between microstructural deformation and macroscopic flow destabilization, $t_R(i) < t_v(i)$, across all monitored locations. At the upstream probe $P_1$ (Fig.~\ref{fig:rg_vs_vel}(a)--(b)), the conformational components reorganize abruptly at $t_R(1) \approx 0.22\lambda$: the normalized streamwise extension $R_{g,xx}/\langle \mathrm{Tr}(\mathbf{R}_g) \rangle$ rises rapidly from its transverse-dominated impingement value ($\sim 0.36$) to $\sim 0.79$, accompanied by a simultaneous drop in $R_{g,yy}/\langle \mathrm{Tr}(\mathbf{R}_g) \rangle$ and the emergence of conformational shear $R_{g,xy}$. Physically, the deviation between $R_{g,xx}$ and $R_{g,yy}$ generates the local first normal stress difference ($\bar{N}_1$), whereas $R_{g,xy}$ drives the shear stress peaks ($\|\bar{\pi}^{xy}_p\|$). This internal chain reorientation occurs well before the kinematic onset time $t_v(1) \approx 0.39\lambda$, when the transverse velocity $v_y$ departs from zero and initiates large-amplitude flapping.

This precursor mechanism extends consistently to the shoulder ($P_2$), shear-layer ($P_3$), and wake ($P_4$) probes in Fig.~\ref{fig:rg_vs_vel}(c)--(d). At each downstream position, the conformational departure at $t_R(i)$ precedes the corresponding velocity destabilization at $t_v(i)$, confirming that the accumulation of elastic energy within the DPD/FENE micro-domains precedes and drives macroscopic wake symmetry breaking. Once the system enters the stationary regime ($t > t_v(i)$), the periodic oscillations of $v_y/\langle \|\mathbf{v}\| \rangle$ (panel c) and the synchronized cycles of $R_{g,xy}/\langle \mathrm{Tr}(\mathbf{R}_g) \rangle$ (panel d) reflect the continuous shedding and transverse flapping of the polymeric tension sheets and sticky flow barriers across $P_2\text{--}P_4$.

\begin{figure}[!ht]
    \centering
    \includegraphics[scale=0.7]{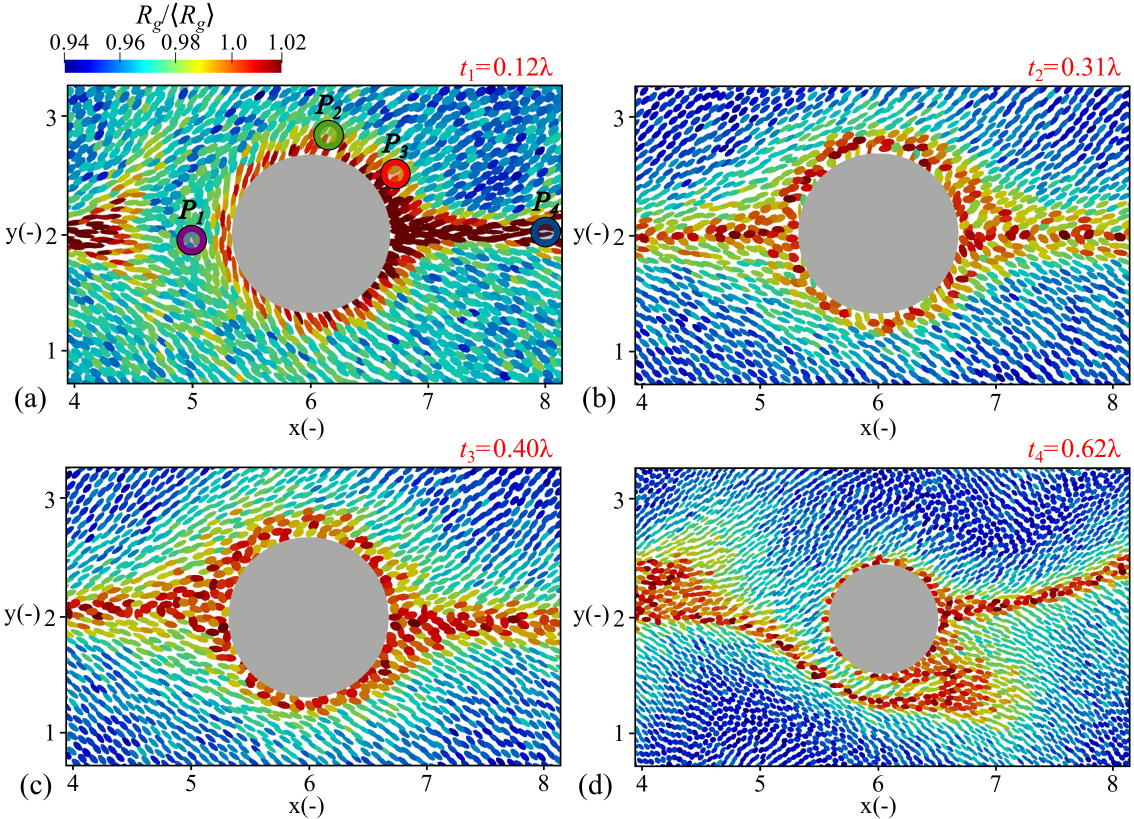}
    \caption{Microstructural zoom around the middle cylinder for $\text{Wi} = 38.8$ ($\text{Re} = 0.015$). (a--d) Close-up snapshots of the 2D tensor glyphs, colored by the normalized scalar radius of gyration $R_g/\langle R_g \rangle$, at $t_1 = 0.12\lambda$, $t_2 = 0.31\lambda$, $t_3 = 0.40\lambda$, and $t_4 = 0.62\lambda$, resolving the formation of the cylinder-wrapping birefringent strand and its subsequent lateral deflection.}
    \label{fig:rg_details_p1}
\end{figure}

Figure~\ref{fig:rg_details_p1} resolves the local glyph kinematics around the middle obstacle and the fine-scale structure of the birefringent strand wrapping the cylinder perimeter, colored by the normalized scalar radius of gyration $R_g/\langle R_g \rangle$ (relative to the instantaneous domain-wide mean $\langle R_g \rangle$) to highlight overall chain extension. During the quasi-steady and pre-kinematic stages ($t_1 = 0.12\lambda$ to $t_3 = 0.40\lambda$), chains impinging near $P_1(5.1, 1.95)$ split around the obstacle into a symmetric envelope of elongated, tilted red ellipsoids that coalesce at the trailing stagnation point ($P_4$). By $t_4 = 0.62\lambda$ (Fig.~\ref{fig:rg_details_p1}(d)), the upstream and trailing birefringent strands deflect asymmetrically downward, channeling incoming macro-particles along a preferential upper pathway. Because the polymer chains are governed by finitely extensible FENE springs and the streamwise extension ($R_{g,xx}$) is partially compensated by transverse compression ($R_{g,yy}$) within the trace $\mathrm{Tr}(\mathbf{R}_g)$, the shift in the scalar mean $R_g$ at $P_1$ remains moderate and does not, by itself, capture the full magnitude of the sub-particle conformational reorganization.

\begin{figure*}[!ht]
    \centering
    \includegraphics[width=\textwidth]{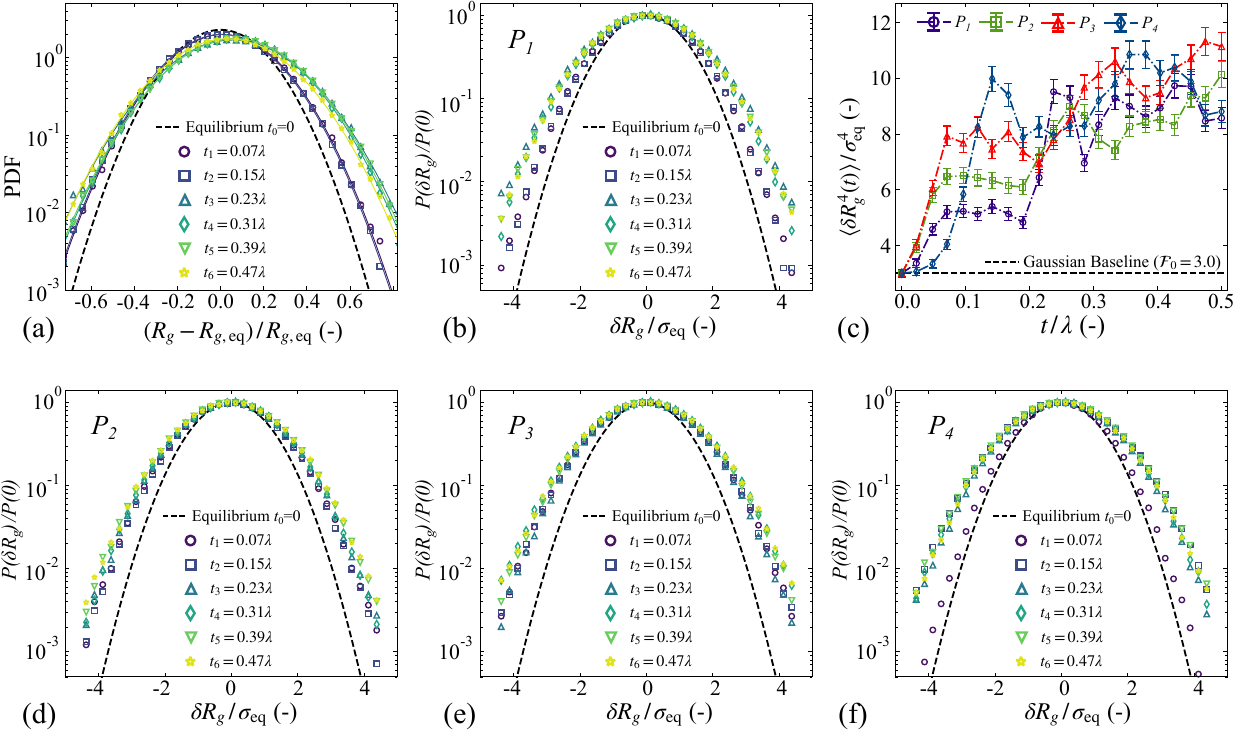}
    \caption{Local evolution of the polymer conformational distribution across probes $P_1\text{--}P_4$, normalized by the quiescent equilibrium baseline ($R_{g,\mathrm{eq}} = 1.076$, $\sigma_{\mathrm{eq}} = 0.188$). (a) Local PDFs of the equilibrium-referenced deviation $(R_g - R_{g,\mathrm{eq}})/R_{g,\mathrm{eq}}$ at $P_1$, showing the bounded shift of the peak alongside progressive widening of the distribution. (b) Locally centered fluctuations $\delta R_g / \sigma_{\mathrm{eq}}$ (with $\delta R_g = R_g - \mu_k(t)$) scaled by the peak value $P(0)$ at $P_1$ (upstream), illustrating local variance growth relative to the equilibrium Gaussian curve (dashed black line). (c) Temporal evolution of the equilibrium-scaled fourth-order central moment $\mathcal{F}(t) = \langle \delta R_g^4(t) \rangle / \sigma_{\mathrm{eq}}^4$ across $P_1\text{--}P_4$ starting from the equilibrium baseline ($\mathcal{F}_0 = 3.0$). (d)--(f) Locally centered distributions $P(\delta R_g)/P(0)$ at (d) $P_2$ (shoulder), (e) $P_3$ (shear layer), and (f) $P_4$ (wake).}
    \label{fig:rg_statistics}
\end{figure*}

To quantify how conformational spread develops around the cylinder, Fig.~\ref{fig:rg_statistics} tracks the local Probability Density Functions (PDFs) of chain extension at the four Eulerian probes ($P_1\text{--}P_4$, Fig.~\ref{fig:rg_details_p1}(a)). At each probe $P_k$, each enclosed SPH particle contributes a local Gaussian micro-domain distribution defined by its instantaneous mean $R_g$ (Eq.~\eqref{eq:rg_scalar_trace}) and sub-particle variance $\sigma_{\mathrm{micro}}^2$ (Eq.~\eqref{eq:micro_variance}); superposing these contributions yields the local probe PDF. Centered fluctuations are defined relative to the instantaneous local probe mean $\mu_k(t)$ as $\delta R_g = R_g - \mu_k(t)$. To compare all four probes on a single, time-independent scale, deviations are normalized by a static quiescent equilibrium baseline ($R_{g,\mathrm{eq}} = 1.076$, $\sigma_{\mathrm{eq}} = 0.188$ in reduced DPD units) obtained from an unstrained micro-DPD simulation.

Figure~\ref{fig:rg_statistics}(a) shows the local conformational PDFs of $(R_g - R_{g,\mathrm{eq}})/R_{g,\mathrm{eq}}$ at the upstream impingement probe $P_1$. At early times ($t_1 = 0.07\lambda$ and $t_2 = 0.15\lambda$), the distribution stays close to the equilibrium curve, with only a small positive shift of the peak and a slight drop in peak density. Past $t_R(1)$ ($t_3 = 0.23\lambda$ to $t_6 = 0.47\lambda$), the profiles separate clearly from equilibrium. Because FENE springs limit finite chain extensibility, the rightward shift of the mean saturates early; the transition therefore appears mainly as a widening of the distribution, most visible on the positive stretching side where mean shift and variance growth combine. This widening reflects the local coexistence of strongly stretched and coiled chains within the same probe volume.

To isolate this local widening from the mean shift, Figs.~\ref{fig:rg_statistics}(b) and \ref{fig:rg_statistics}(d)--(f) center each distribution at its local mean $\mu_k(t)$ and normalize $\delta R_g$ by the fixed equilibrium standard deviation $\sigma_{\mathrm{eq}}$. In this representation, curves that open wider than the dashed equilibrium Gaussian directly reflect an increase in local conformational variance ($\sigma_{\mathrm{loc}} > \sigma_{\mathrm{eq}}$). Collapsing the peaks at the origin reveals a clear spatial sequence around the cylinder: at $P_1$ (Fig.~\ref{fig:rg_statistics}(b)), the profiles remain narrow at $t_1, t_2$ and widen markedly for $t \ge t_3$ once the upstream stagnation zone destabilizes; at the shoulder $P_2$ and shear layer $P_3$ (Figs.~\ref{fig:rg_statistics}(d)--(e)), where wall velocity gradients act immediately, widening appears from $t_1 = 0.07\lambda$ onward; and in the wake $P_4$ (Fig.~\ref{fig:rg_statistics}(f)), the profile stays near equilibrium at $t_1$ due to convective transit time before broadening for $t \ge t_2$ as birefringent strands enter the wake.

Following Lagrangian SPH diagnostics based on PDF broadening and fourth-order moments~\cite{Shi2012}, Fig.~\ref{fig:rg_statistics}(c) summarizes this temporal progression via the equilibrium-scaled fourth-order moment $\mathcal{F}(t) = \langle \delta R_g^4(t) \rangle / \sigma_{\mathrm{eq}}^4$. Scaling by the constant equilibrium variance $\sigma_{\mathrm{eq}}^4$ makes $\mathcal{F}(t)$ grow as $(\sigma_{\mathrm{loc}}(t)/\sigma_{\mathrm{eq}})^4$ from the baseline $\mathcal{F}_0 = 3.0$, amplifying changes in local conformational spread across the array. The four probes activate in sequence: $P_3$ and $P_2$ rise earliest along the cylinder wall, $P_4$ follows after a brief convective delay as the elastic wake forms, and $P_1$ holds an intermediate plateau ($\mathcal{F} \approx 5$) before jumping near $t \approx 0.22\lambda$ once downstream asymmetry feeds back upstream. At later times, all four probes settle in an elevated regime ($\mathcal{F} \approx 8\text{--}11$), confirming that local conformational broadening inside the DPD micro-domains precedes macroscopic symmetry breaking.

\section{Conclusions} \label{sec:conclusions}

This study applies the LHMM to investigate purely elastic instabilities in periodic arrays of cylinders (PAC) at vanishing Reynolds numbers ($\text{Re} \ll 1$). By coupling a macroscopic SPH hydrodynamic solver with microscopic DPD-FENE polymer dynamics, our framework resolves the fine-scale feedback between molecular stretching and flow reconfiguration without relying on empirical constitutive closures. The onset of unsteadiness is marked by spatial symmetry breaking in the transverse velocity field ($v_y$), initiating self-sustained, periodic transverse oscillations. This dynamic is driven by elastic stress accumulation at trailing stagnation zones, where localized bands of high first normal stress difference ($\bar{N}_1$) and shear stress ($\|\bar{\pi}_p^{xy}\|$) propagate downstream. These traveling waves force the main flow to periodically reorient around regions of high elastic resistance, consistent with microfluidic flow visualizations \cite{Haward2020, Hopkins2020, Takizawa2026}.

Bifurcation and spectral analyses quantitatively establish that the transition to unsteadiness occurs via a well-defined onset above a critical Weissenberg number ($\text{Wi}_c \approx 4.0$) and polymer viscosity ratio ($\beta_{p,c} \approx 0.90$). Above these thresholds, downstream advection of unrelaxed elastic stresses between successive obstacles drives a cooperative mixer effect governed by cumulative wake interference \cite{Pan2013, Varshney2018, Browne2020a}. These inter-obstacle wake interactions break single-mode periodic motion and drive the flow into a multi-frequency regime, whose broadband velocity spectra and apparent power-law decay ($\beta \approx 3\text{--}4$) are consistent with elastic-turbulence-like dynamics in obstacle lattices \cite{Berti2008, Chen2025, FoggiRota2026}.

Evaluating spatial confinement and porosity ($\varepsilon$) reveals that increasing the transverse blockage ratio promotes anharmonic, broadband fluctuations even at comparatively modest $\text{Wi}$. In low-porosity media ($\varepsilon \le 0.80$), extreme confinement restricts elastic wakes to narrow channels, pinning stress concentrations within interstitial throats. Conversely, at higher porosity ($\varepsilon = 0.91$), increased spatial clearance allows high-stress filaments, the birefringent strands \cite{Haward2016,Mokhtari2022}, to become mobile and elongated. Our microstructural resolution reveals how these strands act as dynamic stress barriers \cite{Mokhtari2024} that actively reorganize the velocity field into preferential, highly unstable transport pathways \cite{Browne2021, Chen2025}. Consequently, while macroscopic velocity fields become increasingly erratic, the underlying micro-scale stress structures consistently preserve their morphological signatures. Localized stretching--relaxation cycles at stagnation points therefore act as the primary driver of global elastic instabilities in confined porous geometries.

Finally, direct microstructural analysis of the gyration tensor $\mathbf{R}_g$ confirms that these macroscopic stress bands stem from a sharp deviation between longitudinal chain extension ($R_{g,xx}$) and transverse compression ($R_{g,yy}$), coupled with localized conformational shear ($R_{g,xy}$). Time-resolved probe measurements show that microstructural tensor reorganization consistently precedes the onset of macroscopic transverse velocity oscillations ($t_R < t_v$). Furthermore, the local conformational PDFs and the equilibrium-scaled fourth-order moment ($\mathcal{F}$) demonstrate that while the mean chain extension remains bounded by FENE finite extensibility, the destabilization of birefringent strands drives a sharp rise in conformational variance and distribution broadening relative to equilibrium, where strongly stretched and coiled polymer chains coexist within the same micro-domain.

Overall, the LHMM reproduces the main rheological and structural features of creeping viscoelastic instabilities, elastically driven flow reorganization, and broadband spectral decays in periodic obstacle arrays in qualitative agreement with microfluidic observations \cite{Haward2021, Mokhtari2024}. Future work will extend this multiscale formulation to other fully 3D porous networks, where the spatial distribution of obstacles has been shown to strongly influence the onset of elastic instability \cite{Haward2021c, DeBlois2023}, as well as incorporate quantitative comparisons of flow resistance and drag enhancement across varying chain lengths.

\section*{Acknowledgement}
The authors acknowledge the funding provided by IKUR–HPC\&AI – (HPCAI10: MOLD-POLY) and the IKUR Strategy funded by the Basque Government and the European Union NextGenerationEU/PRTR. The research is also partially funded by the Spanish State Research Agency through BCAM Severo Ochoa excellence accreditation CEX2021-0011 42-S/MICIN/AEI/10.13039/501100011033, and through the project PID2024-158994OB-C42 (‘Multiscale Modeling of Friction, Lubrication, and Viscoelasticity in Particle Suspensions’ and acronym ‘MMFLVPS’) funded by MICIU/AEI/10.13039/501100011033 and cofunded by the European Union. The authors gratefully acknowledge the computer resources at MareNostrum and the technical support provided by Barcelona Supercomputing Center (IM-2024-3-0013 and IM-2025-2-0042).

\phantomsection
\section*{Supplementary Information}\label{SM}
Supplementary data associated with this article can be found in the online version. The detailed Supplementary Material, which expands upon the research presented here, including the Supplementary Document and Video, is available temporarily via Mendeley's Dataset repository at: {\color{blue}\url{https://data.mendeley.com/preview/cpx46m36d9?a=c5690eee-7f8c-4ecf-9a7d-9fa7b18ec69e}}. Additionally, the complete code repository for the LHMM framework with DPD-SPH is publicly accessible at: {\color{blue}\url{https://github.com/BCAM-CFD/LHMM/tree/main/LHMM-PolymerDPD}}.


\appendix
\setcounter{figure}{0}
\renewcommand{\thefigure}{\thesection.\arabic{figure}} 

\section{Numerical Convergence and Spectral Analysis}\label{sec:results_1PAC_PSD}
\begin{figure}[!ht]
    \centering
    \includegraphics[width=\columnwidth]{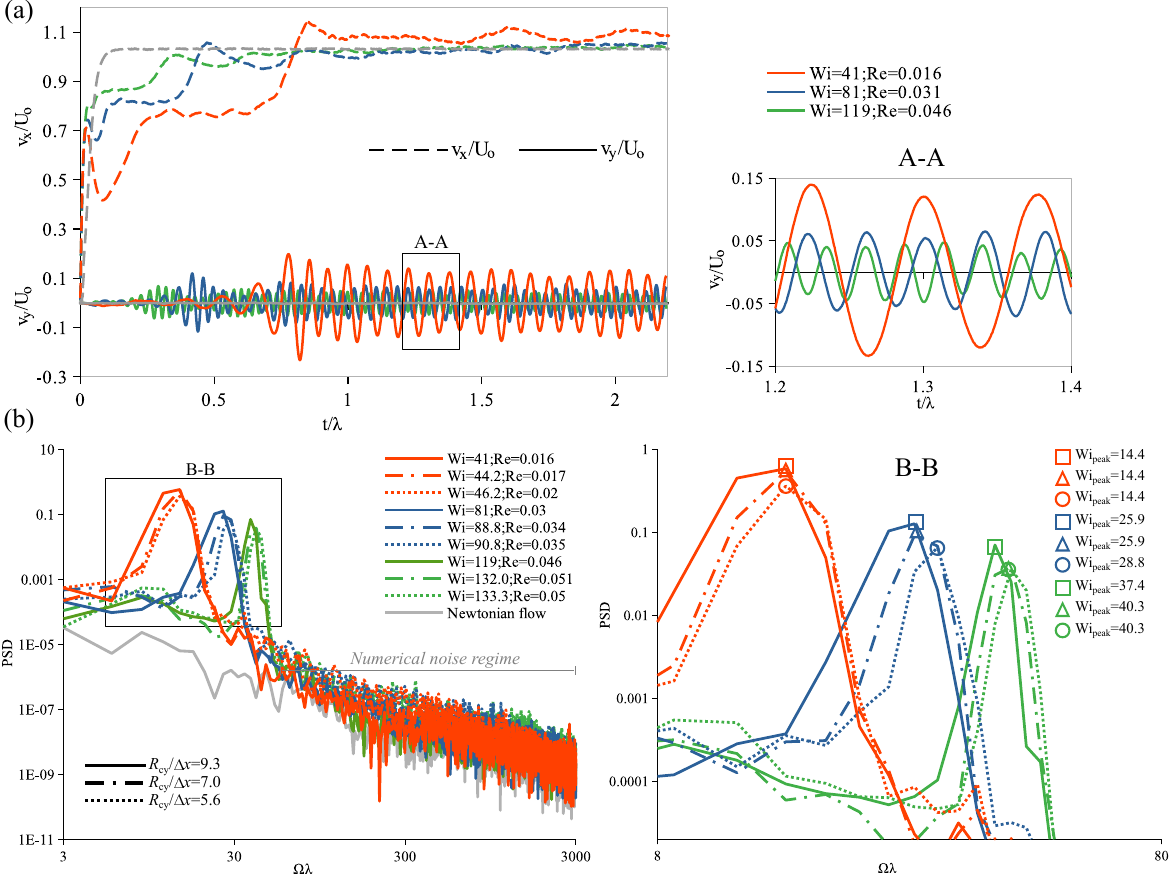} 
    \caption{Temporal evolution, spectral analysis, and numerical convergence of the elastic instability in a single PAC cell for varying $\text{Wi}$, achieved by adjusting $F \in \{0.01, 0.02, 0.03\}$ and $R_{\text{cy}}/\Delta x \in \{5.6, 7.0, 9.3\}$. 
    (a) Evolution of domain-averaged velocity components ($v_x, v_y$) for $\text{Wi} \in \{41, 81, 119\}$ ($\text{Re} \in \{0.016, 0.03, 0.046\}$), with inset A-A detailing the transverse velocity $v_y/U_0$ in the fully developed harmonic regime. 
    (b) PSD of transverse velocity fluctuations across spatial resolutions. The collapse of dominant frequency peaks demonstrates resolution independence. Inset B-B details the peak frequencies defining $\text{Wi}_{\text{peak}} = \Omega_{\text{peak}} \lambda$, highlighted by resolution-specific markers (squares for $R_{\text{cy}}/\Delta x = 9.3$, triangles for $R_{\text{cy}}/\Delta x =7.0$, and circles for $R_{\text{cy}}/\Delta x =5.6$) and color-coded by the body force $F$. Dashed-dotted lines serve as visual guides. The single-cylinder periodic cell ($1\text{L} \times \text{H}$) exhibits a steep spectral drop-off ($\beta \gg 4$) typical of a discrete harmonic limit cycle, contrasting with the broadband, multi-frequency spectra ($\beta \approx 3\text{--}4$) observed in extended arrays.}
    \label{fig:psd_analysisL3}
\end{figure}

The transition of the single-cell PAC from a laminar flow regime to EI with low Reynolds numbers ($\text{Re}\ll1$) is initiated by fundamental spatiotemporal symmetry breaking for $\beta_p=0.97$, $\text{L}=\text{H}=4$, $\text{Wi} \in \{41, 81, 119\}$, and $\text{Re} \in \{0.016, 0.03, 0.046\}$. As observed in Figure~\ref{fig:psd_analysisL3}(a), the streamwise velocity $v_x$ reaches a maximum value following a transient period, deviating from the steady-state Newtonian baseline. The transverse velocity, $v_y$, initially tends toward a steady state before exhibiting a monotonic increase in oscillation amplitude. To quantify the onset of EI, a PSD analysis of the velocity fluctuations was performed. The frequency spectra, presented in Fig.~\ref{fig:psd_analysisL3}(b), were obtained by applying Welch's method with a Hanning window to the dimensionless transverse velocity fluctuations $(v_y - \langle v_y \rangle) / U_0$ during the established unsteady regime. 

Inset A-A of Fig.~\ref{fig:psd_analysisL3}(a) displays a detailed view of this $v_y$ evolution. The temporal signal exhibits a clear harmonic nature, where the oscillation amplitude and frequency are modulated by the $\text{Wi}$ number. This transition from a steady state to a periodic limit cycle, supported by the emergence of a single dominant frequency in the PSD (Fig.~\ref{fig:psd_analysisL3}(b)), is characteristic of a Hopf-like oscillatory instability. In this regime, the nonlinear elastic stresses generated in the micro-domains effectively overcome viscous dissipation to sustain the oscillation. Furthermore, the stationary nature of these limit cycles within a periodic domain identifies them as invariant traveling-wave solutions. This convergence to a coherent, self-sustained state provides numerical evidence of the flow's underlying coherent structures, as the solution preserves its topology while shifting through the periodic lattice, consistent with the framework established for purely EI \cite{Morozov2022, Shnapp2022}.

The frequency spectra in Fig.~\ref{fig:psd_analysisL3}(b) reveal a dominant, narrow-band peak for each $\text{Wi}$ configuration, followed by a regime of numerical noise at high frequency that converges with the case of Newtonian flow. This peak correlates with the characteristic timescales of cyclic polymer stretching and relaxation at the trailing stagnation points \cite{Haward2021,Takizawa2026}. The PSD confirms that in our single PAC unit, the instability manifests as a ``clean'' periodic oscillation where elastic energy is injected and dissipated at a single, well-defined scale, consistent with the localized bistable states described in \cite{Kumar2021a, Varshney2018, Varshney2019a}. To ensure the numerical validity of these results, a spatial resolution study was conducted using $R_{\text{cy}}/\Delta x \in \{9.3, 7.0, 5.6\}$ across the same range of body forces ($F$). As shown in Fig.~\ref{fig:psd_analysisL3}(b), the dominant frequency remains invariant across different discretization scales, demonstrating independence from the particle resolution. Furthermore, the peak Weissenberg number, defined as $\text{Wi}_{\text{peak}} = \Omega_{\text{peak}} \lambda$ (where $\Omega_{\text{peak}}$ represents the fundamental oscillatory frequency), remains consistent across all tested spatial resolutions (Fig.~\ref{fig:psd_analysisL3}(b), inset B-B). The collapse of these spectral peaks demonstrates that the LHMM framework resolves the micro-macro feedback loop, effectively capturing the nonlinear interaction between polymeric memory and macroscopic flow gradients regardless of the particle spacing $\Delta x$. This numerical convergence establishes a reliable particle-based unit-cell solution, which serves as a reference periodic baseline for analyzing how downstream wake interference triggers the mixer effect in larger arrays. 

\subsection{Wake Interference and Spectral Broadening in Multi-Obstacle Arrays}
We explore the reproducibility of this instability by extending the domain to two ($2\text{L} \times \text{H}$) and three ($3\text{L} \times \text{H}$) obstacle configurations with $R_{\text{cy}}/\Delta x\in \{9.3, 7.0\}$. While the underlying physics remains consistent due to periodicity, these extended domains provide a more realistic representation of spatial heterogeneity by reducing the numerical constraints imposed by a single-cell unit. 

\begin{figure}[!ht]
    \centering
    \includegraphics[width=\columnwidth]{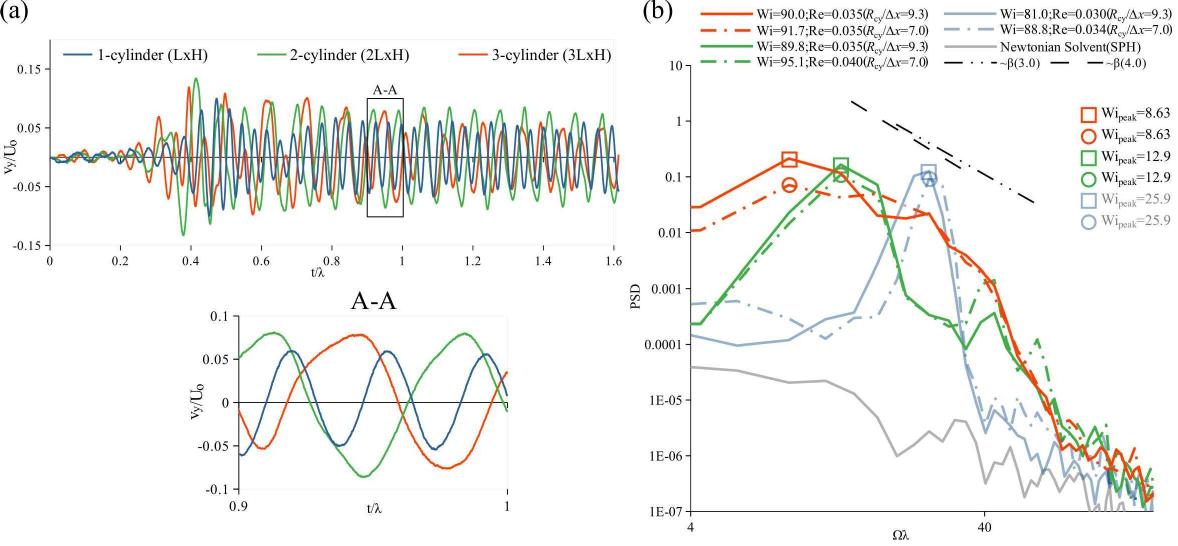}
    \caption{Transition from harmonic oscillations to multi-frequency dynamics in multi-obstacle arrays. (a) Transverse velocity $v_y/U_0$ for 1-, 2-, and 3-cylinder configurations ($R_{\text{cy}}/\Delta x = 7.0$); inset A-A highlights signal distortion as domain length increases. (b) PSD at two spatial resolutions ($R_{\text{cy}}/\Delta x \in \{7.0, 9.3\}$). Markers denote $\text{Wi}_{\text{peak}} = \Omega_{\text{peak}} \lambda$ for each configuration. Visual guides indicate two distinct spectral regimes: a steep harmonic drop-off ($\beta > 4$) characteristic of a low-dimensional limit cycle in single-cylinder periodic domains ($1\text{L} \times \text{H}$), and a broadband spectral decay ($\beta \approx 3\text{--}4$) indicative of multi-scale, elastic-turbulence-like fluctuations in extended arrays ($n \ge 3$).}
    \label{fig:psd_multi}
\end{figure}

The temporal signals for the two- and three-cylinder configurations ($R_{\text{cy}}/\Delta x = 7.0$, Fig.~\ref{fig:psd_multi}(a)) show that the well-defined sinusoidal oscillation of the single-obstacle case undergoes progressive harmonic distortion. The shift of the dominant frequency peak toward lower values alongside the increase in fluctuation amplitude observed in region A-A of Fig.~\ref{fig:psd_multi}(a) indicates that upstream wake interference modifies the self-sustained oscillation. In our PAC lattice, the unrelaxed elastic wake shed by upstream cylinders acts as a finite-amplitude perturbation that destabilizes downstream units, consistent with experimental observations in obstacle arrays \cite{Pan2013, Varshney2018}. This mixer effect promotes the transition toward elastic-turbulence-like spatiotemporal disorder \cite{DeBlois2023}, where interactions between successive wakes populate a continuous range of spatial and temporal scales, as evidenced by the spectral broadening in Fig.~\ref{fig:psd_multi}(b) across resolutions ($R_{\text{cy}}/\Delta x \in \{7.0, 9.3\}$).

The PSD in Fig.~\ref{fig:psd_multi}(b) confirms these multi-scale dynamics across spatial resolutions ($R_{\text{cy}}/\Delta x = 7.0$ and $9.3$). For the single-obstacle benchmark ($1\text{L} \times \text{H}$), elastic energy remains concentrated within a narrow-band harmonic peak centered at a characteristic frequency parameter of $\text{Wi}_{\text{peak}} = 25.9$. As the domain is extended to two ($2\text{L} \times \text{H}$) and three ($3\text{L} \times \text{H}$) obstacles at fixed driving force ($F = 0.02$), the primary peak shifts progressively to $\text{Wi}_{\text{peak}} = 12.9$ and $8.63$, respectively. This inverse scaling ($\text{Wi}_{\text{peak}} \propto 1/n$) reflects the relaxation of the periodic-box wavelength constraint: in single-cylinder ($1\text{L}$) and two-cylinder ($2\text{L}$) periodic domains, periodic images enforce a short fundamental wavelength that locks the wake into high-frequency, narrow-band harmonics, whereas extending the array allows longer-wavelength cooperative wake modes to emerge.

Beyond the shift of the fundamental peak, multi-obstacle configurations exhibit pronounced spectral broadening. Rather than decaying abruptly with a steep slope ($\beta > 4$) typical of single-cylinder periodic cells, the energy redistributes into a broadband spectral decay characterized by effective power-law decay exponents in the range $\beta \approx 3\text{--}4$. This broadening aligns with reported decay slopes for viscoelastic flows past obstacle arrays \cite{Datta2022, Steinberg2021, Singh2024, Shnapp2022}. As shown in Appendix~\ref{sec:scalabilityEI}, for the tested arrays with $n \ge 3$ obstacles, both the spectral envelope and the dominant low-frequency peak become nearly unchanged, indicating that the dominant cumulative wake interactions are captured at this domain length.

\begin{figure}[!ht]
    \centering
    \includegraphics[width=\columnwidth]{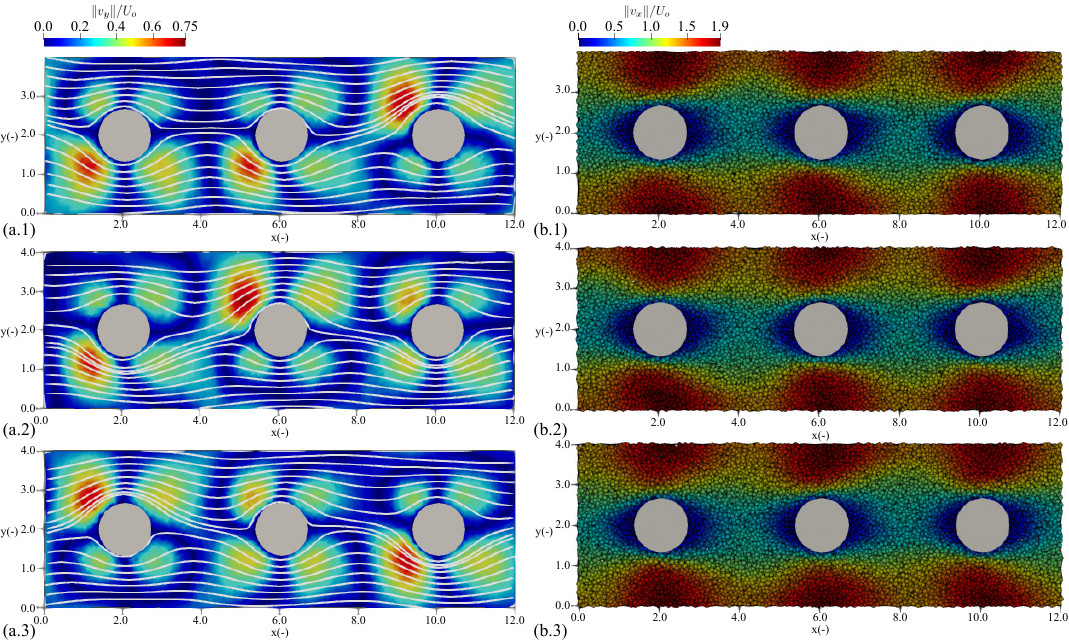}
    \caption{Spatiotemporal evolution of the elastic instability in three-obstacle array configurations ($F=0.02$ and $R_{\text{cy}}/\Delta x = 7.0$). Three sequential snapshots for the $3\text{L} \times \text{H}$ arrangement ($\text{Wi}=91.7$, $\text{Re}=0.035$) at intervals of $\approx 0.5/\Omega_{\text{peak}}$: (a) Streamline patterns colored by normalized transverse velocity $|v_y|/U_0$, showing the periodic whip-like wake motion. (b) SPH particle distribution colored by normalized streamwise velocity $|v_x|/U_0$.}
    \label{fig:multi_obstacle_flow}
\end{figure}

Figure~\ref{fig:multi_obstacle_flow} visualizes the nonlinear oscillatory state in three sequential stages for the $3\text{L} \times \text{H}$ array ($\text{Wi} = 91.7$, $\text{Re} = 0.035$). The transverse velocity component ($|v_y|/U_0$, Fig.~\ref{fig:multi_obstacle_flow}(a)) exhibits significant symmetry-breaking, where the streamlines periodically deviate from the horizontal axis. This lateral deflection and the corresponding wave-like motion reflect the macroscopic manifestation of a phase-locked elastic energy exchange. In contrast, the streamwise velocity component ($|v_x|/U_0$, Fig.~\ref{fig:multi_obstacle_flow}(b)) remains nearly uniform across the main flow direction.

As the domain extends to three obstacles, a transition toward an irregular regime emerges, characterized by a breakdown in spatial coherence, defined here as the degree of phase-synchronization between the oscillations around successive cylinders. The streamlines no longer follow a strictly periodic path and exhibit a spontaneous downward bias past downstream obstacles. This behavior suggests a competition between characteristic frequencies, where persistent high-stress regions, namely the birefringent strands \cite{Mokhtari2024} created by leading cylinders, act as dynamic flow barriers. These strands prevent the fluid from relaxing to its baseline state before encountering the next obstacle, causing a progressive accumulation of elastic stress across the lattice.

\section{Domain Scalability and Dimensional (\texorpdfstring{$n \text{L} \times m \text{H}$}{nL x mH}) Influence on EI}\label{sec:scalabilityEI}
 To evaluate the effect of spatial confinement on the instability, we extend the analysis to multi-column and multi-row configurations ( $n \text{L} \times m \text{H}$). By varying domain dimensions in both streamwise ($x$) and transverse ($y$) directions, we investigate how cumulative elastic stress scales with geometric complexity, allowing us to decouple streamwise wake interference from transverse degrees of freedom.
  
\begin{figure}[!ht]
    \centering
    \includegraphics[width=\columnwidth]{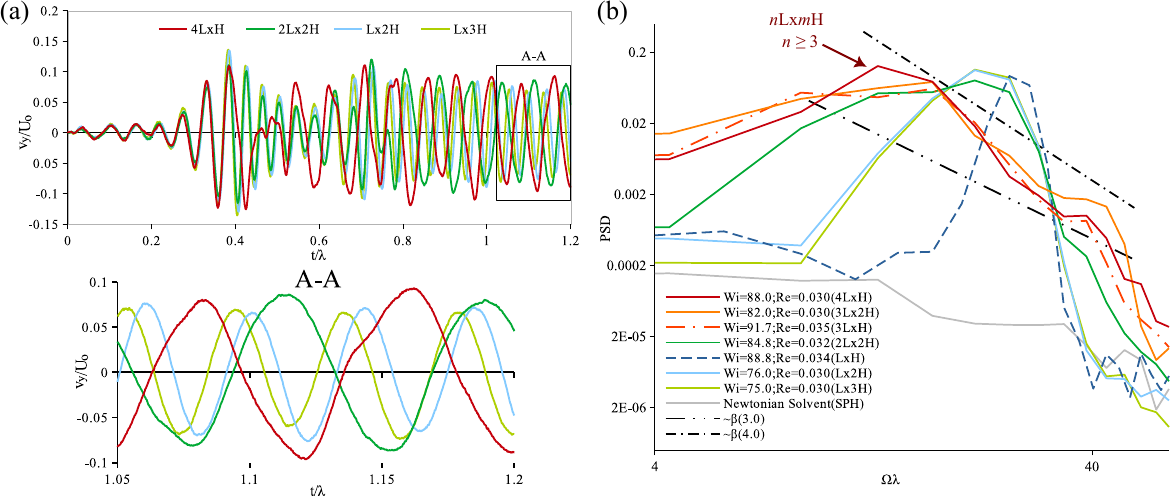}
    \caption{Spatiotemporal sensitivity to domain size. (a) $v_y$ evolution for $n\text{L} \times m\text{H}$ geometries, with inset A-A highlighting harmonic signal distortion. (b) PSD of transverse velocity fluctuations comparing single-unit ($\text{L} \times \text{H}$), longitudinal ($ 3\text{L}, 4\text{L} \times \text{H}$), and multi-row ($2\text{L}, 3\text{L} \times 2\text{H}$) arrays. For the tested arrays with $n \geq 3$, the spectral response becomes nearly unchanged, indicating that the dominant wake-interaction dynamics are captured as the flow transitions from a narrow-band peak to a broadband spectral decay ($\beta \approx 3\text{--}4$).}
    \label{fig:psd_scalabilityL3}
\end{figure}

Figure~\ref{fig:psd_scalabilityL3}(a) displays the transverse velocity $v_y$ for configurations including $4\text{L} \times \text{H}$,  $2\text{L} \times 2\text{H}$, $\text{L} \times 2\text{H}$, and $\text{L} \times 3\text{H}$. Inset A-A reveals fine-scale temporal distortions during $1.05\lambda < t < 1.2\lambda$, where departures from a harmonic response become pronounced as spatial recurrence increases. Increasing the transverse height ($\text{H}$) while keeping length ($\text{L}$) fixed maintains a stable, low-dimensional limit cycle similar to the single-cell unit (cf. Fig.~\ref{fig:psd_multi}(b)), even as the trajectory explores a broader region of phase space. Conversely, in  $4\text{L}\times \text{H}$ and $2\text{L} \times 2 \text{H}$ arrays, the dynamics transition toward a higher-dimensional regime characterized by a broadened frequency spectrum, reflecting the broadband transition observed in extended longitudinal systems \cite{Steinberg2021, Qin2017}. This spectral broadening is corroborated by the PSD analysis (Fig.~\ref{fig:psd_scalabilityL3}(b)), where the spectral profile evolves from a sharp harmonic drop-off into a broadband decay with apparent slopes $\beta \approx 3\text{--}4$, consistent with temporal scaling exponents reported for elastic turbulence in porous geometries \cite{FoggiRota2026, Steinberg2021}.
  
To verify that these multi-scale dynamics reflect physical behavior and are not merely periodic artifacts of the domain, Figure~\ref{fig:psd_scalabilityL3}(b) compares the PSD at a fixed resolution ($R_{\text{cy}}/\Delta x = 7.0$, $F = 0.02$) across domain sizes. Extended domains allow natural wake-wake interference to develop. Therefore, the spectra for the tested arrays with $n \geq 3$ nearly overlap, indicating that the spectral response changes little beyond this length and that the dominant wake-interaction dynamics are captured. The matching spectral scaling ($\beta \approx 3\text{--}4$) confirms that this multi-frequency regime is a converged feature of the PAC geometry.

\begin{figure}[!ht]
   \centering
   \includegraphics[width=\columnwidth]{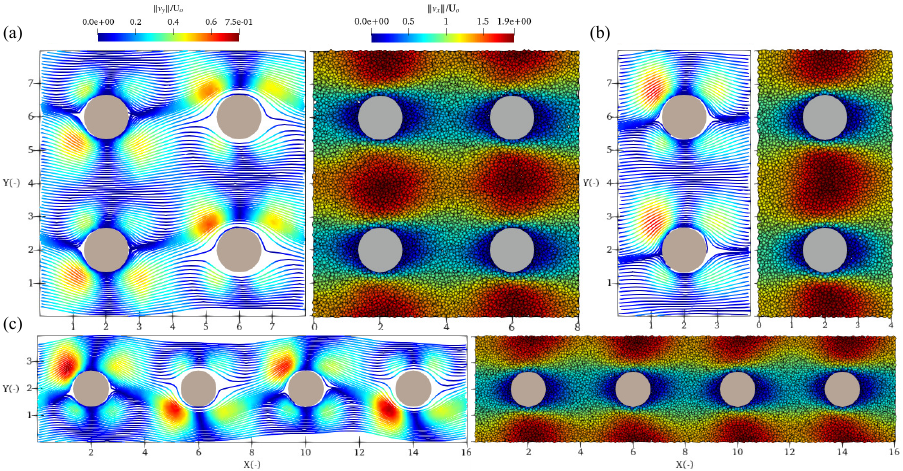}
   \caption{Flow topology in complex PAC geometries ($t > \lambda$). Comparison of (a) $2\text{L} \times 2\text{H}$, (b) $\text{L} \times 2\text{H}$, and (c) $4\text{L} \times \text{H}$. Left panels show streamlines colored by $v_y$ magnitude; right panels show SPH particle distribution colored by $v_x$ magnitude.}
   \label{fig:topology_large}
\end{figure}

The flow topology at $t > \lambda$ (Fig.~\ref{fig:topology_large}) highlights the scalability and stability of the LHMM formulation. Streamlines colored by $v_y$ show that multi-row configurations preserve coherent flow pathways, while the $v_x$ distribution among SPH particles remains consistent even for the $4\text{L} \times \text{H}$ array. This demonstrates the robustness of the LHMM framework in resolving multiscale viscoelastic interactions across extended porous-like lattices, preserving physical flow dynamics without numerical degradation.

\begin{figure}[!ht]
   \centering
   \includegraphics[width=\columnwidth]{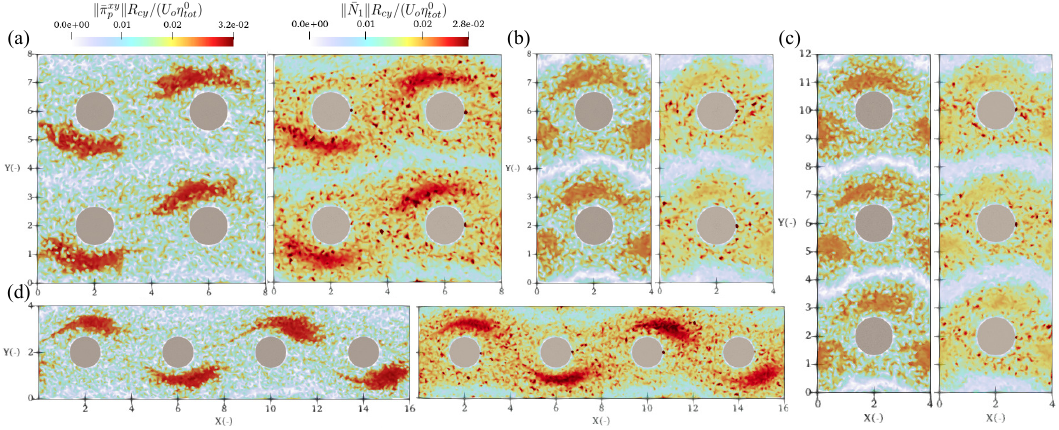}
   \caption{Micro-scale stress fields in the statistically stationary regime ($t > \lambda$). Spatial distributions of $\|\bar{\pi}^{xy}_p\|$ and $\bar{N}_1$ for (a) $2\text{L} \times 2\text{H}$, (b) $\text{L} \times 2\text{H}$, (c) $\text{L} \times 3\text{H}$, and (d) $4\text{L} \times \text{H}$, illustrating the localization of high-stress filaments and sheets across different domain configurations.}
   \label{fig:stress_domain}
\end{figure}

Finally, the spatial distribution of $\|\bar{\pi}^{xy}_p\|$ and $\bar{N}_1$ for the $2\text{L} \times 2\text{H}$, $\text{L} \times 2\text{H}$, $\text{L} \times 3\text{H}$, and $4\text{L} \times \text{H}$ configurations (Fig.~\ref{fig:stress_domain}) confirms the persistence of coherent high-stress structures. High-stress filaments, corresponding to the birefringent strands described by Mokhtari et al. \cite{Mokhtari2022, Mokhtari2024}, remain localized near stagnation points across all array sizes, demonstrating that localized stretching-relaxation cycles are the primary driver of the instability across all tested arrays.

These strands act as dynamic flow barriers that continuously reorganize the local velocity field \cite{Mokhtari2024}. Therefore, this stress-induced barrier mechanism remains robust regardless of domain aspect ratio or macroscopic multi-frequency distortion, confirming that persistent polymer memory at stagnation points drives the nonlinear dynamics in PAC systems regardless of array size \cite{Haward2021c}.

Overall, these domain-size tests show that the mixer effect marks the transition from periodic single-cell limit cycles to multi-obstacle wake interactions. Extending the streamwise domain relaxes the artificial periodicity constraint, lowers the dominant frequency parameter ($\text{Wi}_{\text{peak}}$) to its converged value for $n \ge 3$, and redistributes elastic energy into a broadband spectral decay ($\beta \approx 3\text{--}4$) driven by incomplete polymer relaxation between consecutive obstacles.

\section{Numerical Stability and Scaling Constraints} \label{app:stability_scaling}

Synchronization within the LHMM framework is governed by the temporal constraint:
\begin{equation}
    \delta t = \bar{M} \Delta \bar{t}_{\text{SPH}} = k' \Delta t'_{\text{DPD}}
\end{equation}
To maintain numerical stability, the macroscopic time step $\Delta\bar{t}_{\text{SPH}}$ is reduced upon spatial refinement to comply with the Courant-Friedrichs-Lewy (CFL) condition, as summarized in Table \ref{tab:numerical_params}. The coupling frequency $\bar{M}$ is adjusted proportionally to maintain a fixed synchronization interval of $\delta t = 1.0$. 

Consistent with the rheological identification described in Sec. \ref{sec:micro_results}, the time-scale ratio is set to $\lambda^* = 1$. Under this condition, the dynamical consistency between scales is maintained through the stress scaling factor $\alpha$ (Eq.~\eqref{eq:PI_P}) and the target viscosity ratio $\beta_p$, without requiring additional time-scale corrections.

\begin{table}[ht]
    \centering
    \caption{Numerical parameters for LHMM synchronization ($\delta t = 1.0, k'=200$).}
    \label{tab:numerical_params}
    \begin{tabular}{cccc}
        \hline
        $R_{\text{cy}}/\Delta x$  & $\Delta \bar{t}_{\text{SPH}}$ & $\bar{M}$ & $\delta t$ \\ \hline
        9.3 & 0.001 & 1000 & 1.0 \\
        7.0 & 0.004 & 250  & 1.0 \\
        5.6 & 0.008 & 125  & 1.0 \\ \hline
    \end{tabular}
\end{table}

\section{Detailed DPD/FENE Formulation} \label{app:DPD_details}

The microscopic dynamics of the polymer melt are governed by the DPD framework. Each particle $i$ with mass $m_i$ evolves according to the modified Newton’s equations of motion:
\begin{equation}
m_i \frac{d\mathbf{v}'_i}{dt} = \mathbf{F}_i^{\text{DPD}} + \sum_{j \in \text{bonds}} \mathbf{F}_{ij}^{\text{FENE}},
\label{eq:app_motion}
\end{equation}
where $\mathbf{F}_i^{\text{DPD}}$ represents the sum of non-bonded pairwise interactions:
\begin{equation}
\mathbf{F}_i^{\text{DPD}} = \sum_{j \neq i} \left( \mathbf{F}_{ij}^C + \mathbf{F}_{ij}^D + \mathbf{F}_{ij}^R \right).
\end{equation}

The constituent forces are defined as follows:
\begin{enumerate}
    \item \textbf{Conservative force:} Represents the soft-core repulsion between beads,
    \begin{equation}
    \mathbf{F}_{ij}^C = a_{ij} w^C(r'_{ij}) \mathbf{e}'_{ij},
    \end{equation}
    where $a_{ij}$ is the maximum repulsion and $w^C(r'_{ij}) = (1 - r'_{ij}/r_c)$ for $r'_{ij} < r_c$ (and zero otherwise).
    
    \item \textbf{Dissipative force:} Accounts for viscous resistance,
    \begin{equation}
    \mathbf{F}_{ij}^D = -\gamma^D w^D(r'_{ij}) (\mathbf{e}'_{ij} \cdot \mathbf{v}'_{ij}) \mathbf{e}'_{ij}.
    \end{equation}
    
    \item \textbf{Random force:} Introduces thermal fluctuations,
    \begin{equation}
    \mathbf{F}_{ij}^R = \sigma^R w^R(r'_{ij}) \theta_{ij} \Delta t^{-1/2} \mathbf{e}'_{ij},
    \end{equation}
    where $\theta_{ij}$ is a Gaussian random variable with zero mean and unit variance.
\end{enumerate}

To satisfy the fluctuation-dissipation theorem and ensure the system relaxes to a Gibbs canonical ensemble \cite{Espanol1995}, the following relations are maintained:
\begin{equation}
w^D(r'_{ij}) = [w^R(r'_{ij})]^2 = (1 - r'_{ij}/r_c)^2, \quad \sigma^R = \sqrt{2 \gamma^D k_B T}.
\end{equation}

Chain connectivity is enforced by the FENE potential, which prevents unphysical chain crossing and ensures finite extensibility:
\begin{equation}
\mathbf{F}_{ij}^{\text{FENE}} = \frac{-H_{\text{FENE}} \mathbf{r}'_{ij}}{1 - (\|\mathbf{r}'_{ij}\|/R_0)^2},
\label{eq:app_fene}
\end{equation}
where $H_{\text{FENE}}$ is the spring constant and $R_0$ is the maximum allowable bond length.

\paragraph{Dimensionless Units and Parameters.} 
All simulations are performed in reduced DPD units, where the bead mass ($m_{\rm DPD}$), cutoff radius ($r_c$), and thermal energy ($k_B T$) are set to unity. This defines the characteristic time scale $\tau_{\rm DPD} = r_c \sqrt{m_{\rm DPD}/(k_B T)}$ and stress scale $\pi_{\rm DPD} = k_B T / r_c^3$. Table \ref{Tab:1} summarizes the specific parameters employed for all microsystems in this study.

\section{Micro-to-Macro Coupling}\label{sub_micro_IK}
The Irving--Kirkwood (IK) formalism transfers microscopic stress information from the DPD/FENE subdomains to the macroscale SPH particles \cite{Moreno2021,moreno2023}. It provides a direct estimate of the polymeric stress tensor ($\langle\boldsymbol{\pi}'_{p}\rangle$) that enters the SPH momentum equation (Eq.~\eqref{eq:sph_momentum}). This process is integrated over the temporal interval $\delta t$, capturing the microstructural response of the polymer melt. The formalism maps statistical averages of microscopic quantities (such as particle positions, velocities, and interparticle forces) within the microscale core region ($\Omega'_{\rm core}$) to a representative contribution in the macroscopic stress tensor ($\bar{\boldsymbol{\pi}}_p$).

For a given SPH macro-particle $I$, its associated polymeric stress tensor $\langle\boldsymbol{\pi}'_{p}\rangle_I$ is obtained by microscopically averaging the DPD/FENE interactions within its corresponding micro-domain, $\langle\boldsymbol{\pi}'_{p}\rangle_I = \langle\boldsymbol{\pi}'^{\rm IK}_{p}\rangle_I$. This local stress tensor within $\Omega'_{\rm core}$ is decomposed into kinetic and potential contributions as:
\begin{equation} \label{eq:IKtotal}
    \langle\boldsymbol{\pi}'^{\rm IK}_{p}\rangle_I(\mathbf{x}',t) = \langle\boldsymbol{\pi}'^{\rm kin}_{p}\rangle_I(\mathbf{x}',t) + \langle\boldsymbol{\pi}'^{\rm pot}_{p}\rangle_I(\mathbf{x}',t).
\end{equation}
The kinetic stress ($\langle\boldsymbol{\pi}'^{\rm kin}_{p}\rangle_I$) accounts for momentum transport resulting from the thermal motion and flow-induced fluctuations of individual DPD beads. It captures the energetic and diffusive aspects of the fluid's response to deformation, and it is computed as:
\begin{equation} \label{eq:kinetic_stress}
\langle\boldsymbol{\pi}'^{\rm kin}_{p}\rangle_I = -\frac{1}{N'_t} \sum_{n=1}^{N'_t} \left[
\sum_i m_i\, w_{\text{IK}}(\mathbf{r}'_i(n)-\mathbf{x}')\, \Delta\mathbf{v}'_i(n) \otimes \Delta\mathbf{v}'_i(n)
\right]
\end{equation}
where $N'_t$ is the number of microscopic time steps used in the interval $\delta t' = k' \Delta t_{\rm DPD}$. The quantity $\Delta\mathbf{v}'_i(n)$ is the velocity of particle $i$ relative to the local streaming velocity at time step $n$, and $w_{\text{IK}}(\mathbf{r}'_i(n)-\mathbf{x}')$ is a spatial weighting function that projects microscopic quantities within the sampling volume centered at $\mathbf{x}'$.

The potential stress ($\langle\boldsymbol{\pi}'^{\rm pot}_{p}\rangle_I$) arises from interparticle forces. This includes the FENE potential, which governs the elasticity and finite extensibility of polymer bonds. It also accounts for non-bonded stochastic hydrodynamic interactions such as DPD conservative, dissipative, and random forces. These contributions reflect the stored elastic energy and short-range interactions that characterize the melt's viscoelastic behavior. Assuming a central force decomposition where $\mathbf{f}_{ij} = \varphi_{ij} \mathbf{e}_{ij}$, the potential stress is given by:
\begin{equation} \label{eq:potential_stress}
\langle\boldsymbol{\pi}'^{\rm pot}_{p}\rangle_I = \frac{1}{2N'_t} \sum_{n=1}^{N'_t} \left[
\sum_{\substack{i,j \\ i \neq j}} \mathbf{f}_{ij}(n) \otimes \mathbf{r}'_{ij}(n)\, \mathcal{B}(\mathbf{x}'; \mathbf{r}'_i(n), \mathbf{r}'_j(n))
\right]
\end{equation}
The bond function $\mathcal{B}(\mathbf{x}'; \mathbf{u}', \mathbf{v}')$ integrates the weighting function $w_{\text{IK}}$ along the path connecting the two particles:
\begin{equation}
\mathcal{B}(\mathbf{x}'; \mathbf{u}', \mathbf{v}') = \int_0^1 w_{\text{IK}}\left((1-s)\mathbf{u}' + s\mathbf{v}' - \mathbf{x}'\right) ds
\end{equation}
This formulation yields the Hardy stress. Different choices of the weighting function $w_{\text{IK}}$ recover other well-known microscopic stress definitions. For more details, we refer the reader to the work of Moreno \& Ellero (2021)~\cite{Moreno2021}.

\section{Microscale Domain and Boundary Conditions} \label{sec:boundary_regions}
For all cases, the polymer melt system considered here consists solely of flexible, monodisperse bead-spring chains with $N_b = 16$ beads connected by FENE springs. As polymer melts, these systems contain no solvent particles, and chains are distinguished solely by their bead numbers. The computational domain for these microscale DPD simulations, defining the individual micro-domains within the LHMM framework, is configured using DPD units as illustrated in Fig.~\ref{Fig:1}(b). This domain implements an AFBC to facilitate the coupling between the macroscopic and microscopic scales \cite{Moreno2021,moreno2023}. Table \ref{Tab:2} outlines the dimensions of the core, buffer, and boundary condition regions specific to this setup.

\begin{table}[H]
    \centering
    \caption{Dimensions of the DPD simulation regions for Arbitrary Flow Boundary Conditions (AFBC) used with monodisperse bead-spring chains. $l'_{\text{core}}$ is the length of the core region, $l'_{\text{bc}}$ is the length of the boundary condition region, $l'_{\text{buffer}}$ is the length of the buffer region, and $l'_{\text{total}}$ is the total domain length. For more details, see Fig.~\ref{Fig:1}(b).}
    \begin{tabular}{cccc}
    \hline 
    $l'_{\text{core}}$ & $l'_{\text{bc}}$ & $l'_{\text{buffer}}$ & $l'_{\text{total}}$ \\
    \hline 
    24  & 13  & 13  & 76 \\
    \hline 
    \end{tabular}
    \label{Tab:2}
\end{table}

As illustrated in Fig.~\ref{Fig:1}(b), this computational domain employs three specialized regions to model complex flow configurations and enable the micro-macro coupling:
\begin{itemize}
    \item The \textit{Core region} ($\Omega'_{\rm core}$) represents the active simulation area where the fluid's physical behavior is simulated and where the polymer melt's viscometric properties are measured. Specifically, the average polymer stress $\langle\boldsymbol{\pi}'_p\rangle$ is computed through the IK formalism (Eq.~\eqref{eq:IKtotal}). As shown in Fig.~\ref{Fig:1}(b), this panel displays the initial configuration within $\Omega'_{\rm core}$ for polymer melts with $N_b = 16$. 
    \item The \textit{Boundary-Condition region} ($\Omega'_{\rm bc}$) handles the top-down coupling. It imposes target velocity fields through interpolation functions, specifically utilizing the transfer of discrete macroscopic velocity gradients ($\nabla \mathbf{v}_I$) from the associated macro-particle (Eq.~\eqref{eq:macro_micro}). This region ensures smooth velocity transitions and allows for the specification of arbitrary flow profiles, including both open and closed boundary configurations.
    \item The \textit{Buffer region} ($\Omega'_0$) is designed to stabilize mass and momentum fluxes between periodic boundaries and the core domain. This region effectively acts as a thermal reservoir, with its capacity primarily controlled by adjusting its size ($l'_{\text{buffer}}$) or modulating its temperature \cite{Moreno2021}. While larger buffer regions ensure greater stability at increased computational cost, temperature modulation offers an alternative for maintaining equilibrium without requiring an excessive number of particles. This buffering handles strong velocity gradients without disrupting the core dynamics.
\end{itemize}

\bibliography{library}

\end{document}